\documentclass[aps,superscriptaddress,showpacs,showkeys, reprint, amsmath, amssymb]{revtex4-2}
 
\usepackage{graphicx}
\usepackage{dcolumn}
\usepackage{bm}
\usepackage[colorlinks,linkcolor=blue,citecolor=blue]{hyperref}
\usepackage{subfigure}

\graphicspath{{figs/}}

\begin{document}

\title{Quantum Geometry of Finite XY Chains: A Comparison of Neveu-Schwarz and Ramond Sectors}

\author {N. Einali Saghavaz}
\affiliation{Department of Physics, University of Mohaghegh Ardabili, P.O. Box 179, Ardabil, Iran}
\author {H. Mohammadzadeh}
\affiliation{Department of Physics, University of Mohaghegh Ardabili, P.O. Box 179, Ardabil, Iran}
\author {V. Adami}
\affiliation{Department of Physics, University of Mohaghegh Ardabili, P.O. Box 179, Ardabil, Iran}
\author {M. N. Najafi}
\email{morteza.nattagh@gmail.com }
\affiliation{Department of Physics, University of Mohaghegh Ardabili, P.O. Box 179, Ardabil, Iran}


\begin{abstract}
 This paper presents a geometrical analysis of finite-length XY quantum chains. We begin by examining the ground state and the first excited state of the model, emphasizing the impact of finite-size effects under two distinct choices of the Jordan-Wigner transformation: the Neveu-Schwarz (NS) and Ramond (R) sectors. We explore the geometric features of the system by analyzing the quantum (Berry) curvature derived from the Fubini–Study metric, which is intimately connected to the quantum Fisher information. This approach uncovers a rich interplay between boundary conditions and quantum geometry. In the $\gamma-h$ parameter space, we identify distinct sign-changing arcs of the curvature, confined to the region 
$\gamma^2+h^2<1$. These arcs mark transitions between the NS and R sectors, indicating fundamental changes in the structure of the fermionic ground state. Remarkably, the number of such transition lines increases with system size, hinting at an emergent continuum of topological boundary effects in the thermodynamic limit. Our findings highlight a novel mechanism where boundary conditions shape quantum geometric properties, offering new insights into finite-size topology and the structure of low-dimensional quantum systems. 
\end{abstract}

\maketitle
\tableofcontents

\section{INTRODUCTION}\label{1}
The quantum geometry of a many-body system, encoded in the Fubini–Study metric on the manifold of ground states and the geometric (Berry) phase, captures the holonomy associated with adiabatic parameter evolution and encodes information about underlying phase structures ~\cite{provost1980riemannian,ma2010abelian}. Remarkably, the real part of this metric is directly proportional to the quantum Fisher information, establishing a deep connection between differential geometry and the theory of quantum estimation~\cite{paris2009quantum}. This relationship not only endows geometric quantities with operational meaning but also enables the use of information-theoretic tools to probe criticality and topological features in quantum systems~\cite{zanardi2007bures}. The scalar curvature derived from the metric further offers a glimpse into the curvature of the manifold~\cite{xiao2010berry,haldane2017nobel}, and the fidelity susceptibility, which measures how quickly the system's ground state changes direction in Hilbert space when one tweaks a slow parameter~\cite{gu2010fidelity}. These connections help to identify quantum phase transitions and topological transitions~\cite{haldane2017nobel,chiu2016classification}, as well as the quantum entanglement~\cite{zanardi2007information,you2007fidelity,campos2007quantum,rajabpour2017multipartite}.\\

Quantum spin chains stand out as an exceptional arena for investigating the fundamental and geometrical aspects of quantum many-body physics. These systems, particularly the XY model due to its amenability to analytical treatment~\cite{mbeng2024quantum,zhu2008geometric}, are ideal settings for examining the behavior of quantum states near criticality, and showcase a rich tapestry of geometric and topological features often absent in their higher-dimensional counterparts~\cite{li2024quantum,zhu2008geometric,malakar2024quantum}. The link between geometric phases and critical phenomena not only facilitates the identification of critical regions but also deepens our understanding of the universal features underlying quantum phase transitions~\cite{carollo2005geometric, erdem2019thermodynamic}.\\

Finite systems often exhibit precursors to topological order that appear before the bulk limit is reached, and exhibit unique behaviors that are influenced by boundary effects~\cite{greiner2002quantum, simon2011quantum}. The study of finite-size systems is crucial, especially in light of the increasing experimental realization of quantum systems that are inherently finite, such as in quantum computing platforms~\cite{bose2007quantum} and other condensed matter systems. While extensive research has illuminated the (local and global) properties of XY chains in the thermodynamic limit, finite-size systems have received comparatively less attention~\cite{henkel1987finite,zhu2006scaling}. The role of boundary conditions in shaping the spectral structure and geometric properties remains an open question. The distinction between Neveu-Schwarz (NS) and Ramond (R) sectors stemming from choices in boundary conditions associated with the Jordan-Wigner transformation presents a promising avenue to explore such effects, particularly in relation to the ground state and the excited states, which have been relatively underexplored in the literature~\cite{najafi2020formation}. In conformal field theories, these sectors give rise to different algebraic structures namely, the NS and R superconformal algebras which play a fundamental role in the representation theory of supersymmetric systems~\cite{mavromatos2003neveu,lee2022neveu}. Although a huge amount of studies have extensively analyzed the algebraic representation spaces of these sectors (particularly focusing on unitarity, modular invariance, and highest-weight modules), the physical implications of these mathematical structures in finite quantum spin systems, such as the XY chain, remain less explored~\cite{sakurai2020modern,bigan2020quantum,coraggio2025quantum,carollo2005geometric,zhu2006scaling}. In particular, their impact on the spectral geometry, topological features, and thermodynamic behavior of finite systems has not been systematically investigated.

This paper sets out to investigate the quantum (local and geometrical) properties of finite XY chains in terms of the anisotropy parameter $\gamma$ and the transverse magnetic field $h$ in both NS and R sectors. By analyzing the spectrum of energy eigenstates, we aim to unravel how each sector influences the system's geometry, seeking to fill a void in the existing literature concerning their role in finite-size corrections and metric behavior~\cite{cohn1986covariant}. A parametric analysis of the quantum geometry encompassing both ground and excited states will be carried out to elucidate the behavior of the Fubini-Study metric and the geometric phase in parameter space. Our investigation reveals that in the $\gamma-h$ phase space, the Berry curvature exhibits distinct sign changes along a set of well-defined curves within the ordered phase, specifically in the region $\gamma^2+h^2<1$. These curves are associated with transitions between the NS and R sectors, reflecting a fundamental change in the fermionic ground state structure. Unlike conventional quantum phase transitions driven by bulk gap closures, the observed transitions are rooted in the interplay between topology and finite-size boundary conditions. This behavior underscores the geometric sensitivity of the Berry curvature to the underlying boundary conditions, providing a new perspective on topological phase structures and the role of boundary effects in defining the quantum geometry in low-dimensional quantum spin systems.

This paper is structured as follows: Section~\ref{SEC:Aims} is devoted to introducing the aims and the achievements of the present paper. In Section~\ref{2}, we will revisit the fermionic representation of the XY chain. Section~\ref{4} will introduce the framework of quantum geometry. Section~\ref{3} will present our analytical and numerical findings for the NS and R sectors. In Section~\ref{SEC:Geometry}, we will provide a comparative geometric analysis. Finally, Section~\ref{7} will offer concluding remarks and outline directions for future work.

\section{AIMS AND ACHIEVEMENTS}\label{SEC:Aims}
In this paper, we develop a unified treatment for analyzing the effect of finite sizes in observables in spin chains. Here we summarize the findings of this paper. After exploring some symmetry identities for spin chains (SEC~\ref{SEC:symmetry} and summarized in Table~\ref{tab:conditions}) and exploring Jordan Wigner transformation (SEC~\ref{SEC:JWMain}), we identify the energy spectrum in various regions that are discussed in SEC.~\ref{SEC:definitions}. The findings of the paper are as follows: 
\begin{itemize}
    \item We explore the energy spectrum in the $\gamma-h$ space, and identify the energy difference of NS and R sectors in SEC.~\ref{SEC:EnergySpec}. This difference behaves differently according to Eqs.~\ref{Eq:exponential} and~\ref{Eq:power-Law}, which are plotted and analyzed in Figs.~\ref{Fig:PhaseDiagramEnergy} and~\ref{Fig:delta-E}. Equation~\ref{Eq:EnergyDifference} gives the energy difference between NS and R sectors in the critical case. The exponents are plotted and analyzed in Fig.~\ref{fig:dE-alpha}.
    \item We test how the quantum geometric observables evolve upon increasing the size of the system ($L$) in SEC.~\ref{SEC:Geometry}. The results are partially gathered in Figs.~\ref{Fig:curvature_combined} and~\ref{fig:h1}. 
    \item We summarize the $L$ dependence of the scaler curvature associated with the Fubini-Study metric in Eqs.~\ref{Eq:Rx1},~\ref{Eq:ROscillating1},~\ref{Eq:ROscillating2} and~\ref{Eq:Rx2}, where some critical and non-critical exponents are introduced and presented in Figs.~\ref{fig:exponential}. The phase diagrams of NS and R sectors, based on the exponents are sketched in Fig.~\ref{fig:schematic}.
    \item Various regions are identified based on the sign of the scaler curvature which are analyzed in Fig.~\ref{fig:phase_spaces} in terms of $L$, and NS and R sectors. This tells us more about the correlation between the boundary conditions, and the geometry and topology of finite spin chains.
\end{itemize}
 
\section{THE XY CHAIN}\label{2}
We consider spin-$\frac{1}{2}$ quantum XY Model on a one-dimensional chain characterized by anisotropic exchange interactions. The Hamiltonian governing the XY chain is expressed as:

\begin{equation}
\begin{split}
H_{\text{XY}} & = -J \sum_{l=1}^L \left[\frac{1}{4}(1 + \gamma) \sigma_l^x \sigma_{l+1}^x + \frac{1}{4}(1 - \gamma) \sigma_l^y \sigma_{l+1}^y \right] \\
& \quad - \frac{h}{2} \sum_{l=1}^L \sigma_l^z,
\end{split}
\label{Hamiltonian}
\end{equation}
where $\sigma_l^i$ ($i = x, y, z$) are Pauli matrices, $J$ is the exchange interaction strength, $h$ represents the transverse magnetic field, $\gamma$ is the anisotropy parameter, and $L$ is the chain length. We consider periodic boundary conditions for the chain, i.e. $\sigma^a_{L+1}\equiv\sigma^a_1$, $a=x,y,z$. The model captures the interaction between nearest-neighbor spins. This system has gained significant attention due to its intriguing properties, such as integrability, allowing for exact solutions~\cite{franchini2017introduction,najafi2020formation}, and provides a valuable framework for studying geometric phases and quantum phase transitions~\cite{carollo2020geometry}. \\

\subsection{Symmetries}\label{SEC:symmetry}
The XY Hamiltonian has explicit symmetries leading to symmetry identities for the eigen energies. The first example is associated with $x-y$ plane reflection $\sigma_x\to -\sigma_x$, $\sigma_y\to -\sigma_y$, and $\sigma_z\to \sigma_z$ expressed using the transformation $\hat{N}_LH_{\text{XY}}(J,\gamma,h)\hat{N}_L=H_{\text{XY}}(J,\gamma,h)$ where 
\begin{equation}
\hat{N}_L\equiv \prod_{j=1}^L\sigma_j^z.
\label{Eq:N}
\end{equation}
Consider the Ising model $\gamma=1$, for which, the ground state is found to be doubly degenerate at $h=0$:
\begin{equation}
\begin{split}
&\left| \text{gs}_{\gamma=1,h=0}\right\rangle_1=\left|\uparrow\uparrow\uparrow...\uparrow \right\rangle_x \\
& \left| \text{gs}_{\gamma=1,h=0}\right\rangle_2=\hat{N}_L\left| \text{gs}_{\gamma=1,h=0}\right\rangle_1,
\end{split}
\end{equation}
where $\left| s_1s_2...s_L\right\rangle_x\equiv \left|s_1 \right\rangle_x \bigotimes\left|s_2 \right\rangle_x \bigotimes...\left|s_L \right\rangle_x$, and $\left|s \right\rangle_x$ is an $s$ eigenvector of $\sigma_x$. This degeneracy is lifted as $h$ becomes non-zero. This is already expected from the Kramer's theorem since for $h=0$ the system is time reversal symmetric. Generally, the time-reversal symmetry is defined as $\hat{\mathcal{T}}=-i\sigma^yK$ for spin $\frac{1}{2}$ systems, where $K$ is complex conjugation. This transformation leaves the Hamiltonian unchanged only for $h=0$, so that one expects that the system is at least doubly degenerate for $h=0$.
\begin{table*}
	\center
	\caption{A list of some discrete and continuous symmetries. For the definitions and descriptions refer to the text.}
	\label{tab:conditions}
	\begin{tabular}{|c | c | c | }
		\hline transformation & generator & Consequence \\
		\hline $\sigma_l^x\to -\sigma_l^x$, $\sigma_l^y\to -\sigma_l^y$, $\sigma_l^z\to \sigma_l^z$ & $\hat{N}_L\equiv \prod_{j=1}^L\sigma_j^z$ & A double degeneracy \\
        \hline
        $\sigma_l^x\to \sigma_l^x$, $\sigma_l^y\to -\sigma_l^y$, $\sigma_l^z\to -\sigma_l^z$ & $\hat{\mathcal{T}}=\prod_{j=1}^L\left(-i\sigma_j^yK\right)$ & Kramer's degeneracy \\
        \hline
     $\sigma_l^x\to \sigma_l^y$, $\sigma_l^y\to -\sigma_l^x$, $\sigma_l^z\to \sigma_l^z$ & $\hat{P}_\gamma\equiv\prod_{j=1}^L\exp [i\frac{\pi}{4}\sigma_j^z]$ & $E_{J,-\gamma,h}=E_{J,\gamma,h}$ \\
         \hline
          $\sigma_{l}^x\to \sigma_{l}^x$, $\sigma_{l}^y\to -\sigma_{l}^y$, $\sigma_l^z\to -\sigma_l^z$ & $\hat{P}_h\equiv \prod_{l=1}^{L}\sigma_l^x$ & $E_{J,\gamma,-h}=E_{J,\gamma,h}$ \\
         \hline
         $\sigma_{2l-1}^x\to -\sigma_{2l-1}^x$, $\sigma_{2l-1}^y\to -\sigma_{2l-1}^y$, $\sigma_l^z\to \sigma_l^z$ & $\hat{P}_J\equiv \prod_{l=1}^{[\frac{L}{2}]}\sigma_{2l-1}^z$ & $E_{-J,\gamma,h}=E_{J,\gamma,h}$ \\
        \hline
XX model: $(\sigma_{l}^x,\sigma_{l}^y)\to \mathcal{R}_\theta(\sigma_{l}^x,\sigma_{l}^y)$ & $\mathcal{R}_\theta\equiv\prod_{j=1}^L \exp[\frac{i\theta}{2}\sigma_j^z]$ & A continuous symmetry \\
        \hline
	\end{tabular}
	\label{tab:Projections}
\end{table*}
In table~\ref{tab:Projections} we summarize the consequences of the $x-y$ plane reflection $\hat{N}_L$, time reversal symmetry $\hat{\mathcal{T}}$ rotation $\hat{P}_\gamma$, staggered reflection symmetry $\hat{P}_J$ and magnetic field inversion symmetry $\hat{P}_h$, and full rotational symmetry $\mathcal{R}_\theta$ for $\gamma=0$ (XX model). In the rest of the paper we set $J\equiv 1$.

\subsection{Diagonalizing the XY Hamiltonian}\label{SEC:JWMain}
The Jordan-Wigner transformation provides a way to exactly diagonolize the XY hamiltonian which is particularly notable for its exact solvability (see appendix~\ref{SEC:JW} for details on the Jordan-Wigner transformation). This transformation maps the spin operators into fermionic ones in one dimension~\cite{batista2001generalized, lieb1961two}. More precisely, the following non-local transformations map the spin operators to string operators, which are fermionic creation ($\bar{c}^{\dagger}_l$) and annihilation ($\bar{c}_l$) operators respectively:
\begin{equation}
\bar{c}_l^\dagger = \hat{N}_{l-1} \sigma_l^+, \quad \bar{c}_l = \hat{N}_{l-1} \sigma_l^-,
\end{equation}
where $\hat{N}_l$ is given in Eq.~\ref{Eq:N} and is equal to $\prod_{j\le m}\left(2\bar{c}_j^{\dagger}\bar{c}_j-1\right)$ and $\sigma_l^+ = \frac{1}{2} (\sigma^x + i\sigma^y)$ and $\sigma_l^- = \frac{1}{2} (\sigma^x - i\sigma^y)$. Note that the parity, defined in the spin chain as
\begin{equation}
\mathcal{P}\sigma^a_l\mathcal{P}^{-1}=\sigma^a_{L-j+1},\ a=x,y,z,
\end{equation}
operates on the fermionic operators as
\begin{equation}
\mathcal{P}\bar{c}_l\mathcal{P}^{-1}=\hat{N}_L\bar{c}_{L-j+1},
\end{equation}
so that $\hat{N}_L$ determines also the parity of the fermionic chain. This is a reason that $\hat{N}_L$ is also called fermion number parity. The fermionic boundary conditions also read
\begin{equation}
\left(\begin{matrix}
\bar{c}_{L+1} \\
\bar{c}_{L+1}^{\dagger}
\end{matrix}\right)=\hat{N}_{L}\left(\begin{matrix}
\bar{c}_1 \\
\bar{c}_1^{\dagger}
\end{matrix}\right).
\label{Eq:FBC}
\end{equation}
This shows that the fermionic \textit{sectors} are classified in terms of the eigenvalues of $\hat{N}_L$. Showing the eigenvalue of $\hat{N}_L$ by $-N_L$ (the minus sign is a convention), one can easily check that $\hat{N}_L^2=1$, so that its eigenvalues are either $N_L=-1$ or $N_L=+1$, labeled as even (odd) states.\\

As a result of the Jordan-Wigner transformation, the XY Hamiltonian is mapped onto
\begin{equation}
\begin{aligned}
H_{XY}^F &= \frac{J}{2} \sum_{l=1}^{L-1} \left(\bar{c}^\dagger_l\bar{c}_{l+1}+\gamma \bar{c}^\dagger_l\bar{c}^\dagger_{l+1} \right) - \frac{J\hat{N}_{L-1}}{2} \left(\bar{c}^\dagger_L \bar{c}_1+\gamma \bar{c}^\dagger_L \bar{c}^\dagger_1\right) \\
& - \frac{h}{2} \sum_{l=1}^L \bar{c}^\dagger_l \bar{c}_l + \frac{hL}{4} +H.C. ,
\end{aligned}
\label{Eq:HF_XY}
\end{equation}
where $H.C.$ represents hermitian conjugate. This when combined with Eq.~\ref{Eq:FBC} leads to (note that the last term of the first line of Eq.~\ref{Eq:HF_XY} reads $-\hat{N}_{L-1}\bar{c}^{\dagger}_L\bar{c}_1=-\hat{N}_{L-1}\bar{c}^{\dagger}_L\hat{N}_{L}\bar{c}_1$, giving rise to to $\bar{c}^{\dagger}_L\bar{c}_{L+1}$):
\begin{equation}
H_{XY}^F = \frac{J}{2} \sum_{l=1}^L \left(\bar{c}_l^\dagger \bar{c}_{l+1} + \gamma \bar{c}_l^\dagger \bar{c}_{l+1}^\dagger \right) - \frac{h}{2} \sum_{l=1}^L \bar{c}^\dagger_l \bar{c}_l +H.C.
\end{equation}
where we ignored the last unimportant constant. We will work with this Hamiltonian throughout them paper.\\

Since the commutator $[H_{XY}^F,\hat{N}_L]$ is zero, they have simultaneous eigenstates, i.e. the energy eigenstates are labeled using the eigenvalues of $\hat{N}_L$: 
\begin{equation}
\begin{split}
&H^F_{XY} |E_\nu,N_L\rangle = E_\nu |E_\nu,N_L\rangle,\\
&\hat{N}_L |E_\nu,N_L\rangle = -N_L |E_\nu,N_L\rangle,
\end{split}
\end{equation} 
where $\nu$ enumerates the energy spectrum. The odd ($N_L=+1$) and the even ($N_L=-1$) eigenstates correspond to anti-periodic and periodic boundary conditions respectively in the fermionic sector. This identifies the \textit{Neveu-Schwarz} (NS with anti-periodicity) sector labeled by $N_L=+1$ and the \textit{Ramond} (R with periodicity) sector labeled by $N_L=-1$.\\

Defining a new fermionic field as follows
\begin{equation}
c_{l} = \exp\left[i \frac{\pi(N_L+1)}{2L} l\right] \bar{c}_{l},
\end{equation}
retrieves a periodic boundary conditions for both NS and R sectors:
\begin{equation}
\left(\begin{matrix}
c_{L+1} \\
c_{L+1}^{\dagger}
\end{matrix}\right)=\left(\begin{matrix}
c_1 \\
c_1^{\dagger}
\end{matrix}\right),
\end{equation}
leading to a translational invariant Hamiltonian~Eq.~\ref{Eq:H_XY-Transformed} which can analyzed in the Fourier space. While for $N_L=-1$ the transformation is trivial, for $N_L=+1$ this transformation leads to a phase shift for the Fourier components. The Fourier transformation is implemented by 
\begin{equation}
c_{l} = \frac{1}{\sqrt{L}} \sum_{k=1}^L e^{i\zeta_kl}\tilde{c}_{k},\  c^{\dagger}_{l} = \frac{1}{\sqrt{L}} \sum_{k=1}^L e^{-i\zeta_kl}\tilde{c}^{\dagger}_{k}
\end{equation}
where $\zeta_k\equiv \frac{2\pi}{L}k$, and $\tilde{c}_k$ and $\tilde{c}_k^{\dagger}$ are annihilation and creation operator in the Fourier space, satisfying
\begin{equation}
\left\lbrace \tilde{c}^{\dagger}_k,\tilde{c}_{k'}\right\rbrace =\delta_{k,k'}, \ \left\lbrace \tilde{c}^{\dagger}_k,\tilde{c}_{k'}^{\dagger}\right\rbrace = \left\lbrace \tilde{c}_k,\tilde{c}_{k'}\right\rbrace =0.
\end{equation}

The Hamiltonian can be rewritten in terms of new operators $\eta_k^T = (\tilde{c}_k,\tilde{c}_{-k + \frac{N_L+1}{2}}^{\dagger})$ and $\eta_k^{\dagger} = (\tilde{c}_k^{\dagger},\tilde{c}_{-k + \frac{N_L+1}{2}})$ as
\begin{equation}
H_{XY}^F = \frac{1}{2} \sum_{k} \eta_k^\dagger \mathcal{H}_k \eta_k,
\end{equation}
where $\mathcal{H}_k$ is a  matrix given by:
\begin{equation}
\mathcal{H}_k = 
\begin{pmatrix}
J \cos\phi^{(N_L)}_k - h & i\gamma J \sin\phi^{(N_L)}_k \\
-i\gamma J \sin\phi^{(N_L)}_k & -J \cos\phi^{(N_L)}_k + h
\end{pmatrix}.
\label{Eq:HamilMatrix}
\end{equation}
where $\phi^{(N_L)}_k\equiv \zeta_k-\frac{\pi(N_L+1)}{2L}$. One can diagonalize the Hamiltonian using the Bogoliobov transformation 
\begin{equation}
\begin{split}
&\mathcal{B}_k\equiv e^{\frac{-i}{2}\theta^{(N_L)}_k\sigma^x}\eta_k\equiv\left(\begin{matrix}
b_k\\
b^{\dagger}_{-k+\frac{N_L+1}{2}}
\end{matrix}\right),\\  &\mathcal{B}_k^{\dagger}\equiv e^{\frac{i}{2}\theta^{(N_L)}_k\sigma^x}\eta_k^{\dagger}=\left(b_k^{\dagger},
b_{-k+\frac{N_L+1}{2}}\right),
\end{split}
\end{equation}
where $\theta^{(N_L)}_{k} \equiv \tan^{-1}\left(\frac{\gamma \sin\phi^{(N_L)}_k}{h-\cos\phi^{(N_L)}_k}\right)$, and $b_k$ and $b^{\dagger}_{k}$ are the corresponding components. This transformation gives rise to the following diagonalized form
\begin{equation}
H_{XY}^F=\frac{1}{2}\sum_k\epsilon_{N_L}(k)\mathcal{B}_k^{\dagger}\sigma_z\mathcal{B}_k,
\label{Eq:HDiag}
\end{equation}
where 
\begin{equation}
\epsilon_{N_L}(k) \equiv \sqrt{\left(h-\cos\phi^{(N_L)}_{k}\right)^2 + \gamma^2\sin^2\phi^{(N_L)}_{k}},
\label{Eq:SPEnergy}
\end{equation} 
is the single particle excitation energies. Equation~\ref{Eq:HDiag} casts finally to the following diagonalized form
\begin{equation}
\begin{split}
H_{XY}^F=\sum_k\epsilon_{N_L}(k)\left(b_k^{\dagger}b_k-\frac{1}{2}\right),
\end{split}
\label{Eq:HXYDiag}
\end{equation}
the spectrum of which depends on the selected sector, which is hidden in the dependence on $N_L$: $\phi^{(-1)}_k=\zeta_k$ for R sector, while $\phi^{(1)}_k=\zeta_k-\frac{\pi}{L}$ for the NS sector, see Fig.~\ref{Fig:App:R-and-NS} in Appendix~\ref{SEC:App-diagonalization}.

\subsection{The Ground and First Excited States}

To construct the ground state, we find the iso-spin state in the $k$th mode. The ground state is a product state of the models with lowest available anergies:
\begin{equation}
|\text{gs},N_L\rangle = \bigotimes_{k}\left| \downarrow\right\rangle_k^{N_L} 
\label{Eq:GSMain}
\end{equation}
where
\begin{equation}
\left| \downarrow\right\rangle_k^{N_L} \equiv{\mathcal{B}_k^{(2)}}^{\dagger}\left| 0\right\rangle=\left(\begin{matrix}
i\sin\frac{\theta_k^{N_L}}{2}\\
\cos\frac{\theta_k^{N_L}}{2}
\end{matrix}\right),
\end{equation}
as can be confirmed either directly by diagonalization of Eq.~\ref{Eq:HamilMatrix}, or using the Bogoliobov transformation leading to Eq.~\ref{Eq:AppGS}. In this relation $\left| \downarrow\right\rangle_k^{N_L}$ is a downward isospin state, and should not be confused with real spins of the XY model. A local excitation in $k$th mode is generated by flipping the isospin as follows:
\begin{equation}
\left| \uparrow\right\rangle_k^{N_L} \equiv{\mathcal{B}_k^{(1)}}^{\dagger}\left| 0\right\rangle\propto\left(\begin{matrix}
-i\cos\frac{\theta_k^{N_L}}{2}\\
\sin\frac{\theta_k^{N_L}}{2}
\end{matrix}\right),
\end{equation}
These two states are related by the relation
\begin{equation}
\partial_x\left| \downarrow\right\rangle_k^{N_L}=-\frac{\partial_x\theta^{N_L}_k}{2}\left| \uparrow\right\rangle_k^{N_L} ,
\end{equation}
where $x=h,\gamma$. Therefore, a partial derivative of the ground state leads to the following identity
\begin{equation}
\begin{split}
\partial_x|\text{gs},N_L\rangle &= \sum_{k}\left(\bigotimes_{k'\ne k}\left| \downarrow\right\rangle_{k'}^{N_L} \right)\partial_x\left| \downarrow\right\rangle_k^{N_L} \\
&=-\sum_{k}\left(\bigotimes_{k'\ne k}\left| \downarrow\right\rangle_{k'}^{N_L} \right)\left(\frac{\partial_x\theta^{N_L}_{k}}{2}\right)\left| \uparrow\right\rangle_k^{N_L}  \\
&= \sum_{k}a_{k}^{(x)}|e,N_L\rangle_{k},
\end{split}
\end{equation}
where
\begin{equation}
|e,N_L\rangle_{k} \equiv \bigotimes_{k'}|s,N_L\rangle_{k',k},\quad a_k^{(x)} \equiv -\frac{\partial_x\theta^{N_L}_{k}}{2},
\end{equation}
and
\begin{equation}
|s,N_L\rangle_{k',k}=   
\begin{cases}
\left| \downarrow\right\rangle_{k'}^{N_L}  & \text{if}\ k\ne k' \\
\left| \uparrow\right\rangle_k^{N_L}  & \text{if}\ k= k'
\end{cases}.
\end{equation}
The first excited state of the system is then found to be
\begin{equation}
\left|\text{FES},N_L \right\rangle =\text{min}_{E(\left| e,N_L\right\rangle_k)}\left\lbrace \left| e,N_L\right\rangle_k | k\in[1,L]   \right\rbrace .
\label{Eq:FES}
\end{equation}
Noting that $|e,N_L\rangle_k$ is an excited state corresponding to the excitation of a fermion in the mode $k$, this relation affirms that the partial derivative of the ground state is a linear combination of the excited states. This relation helps in calculating the inner product of the quantum states with $\partial_x|gs\rangle$, which is required for calculating the Berry's tensor elements in the following sections.

\subsection{The definition of the regions}\label{SEC:definitions}

The results of this paper are classified in terms of the regions in the $\gamma-h$ space. There are two critical regions for the XY model in the thermodynamic limit ($L \to \infty$) where the ground-state energy gap closes, namely $\gamma=0$ ($-1 \leq h \leq 1$) and the line $h = \pm 1$, while for the other regions the properties are different. On the line $\gamma^2+h^2=1$ the ground state of the finite spin chain becomes degenerate, and is equal to the ground state of the NS and R fermionic chain simultaneously. This line is recognized as the parity transition point~\cite{franchini2007renyi}, on which the ground state becomes doubly degenerate. More precisely, the ground state can be written as the following product states~\cite{muller1985implications}:
\begin{equation}
\begin{split}
&\left| \text{gs}\right\rangle_1=\prod_{l}\left(\cos\chi\left|\uparrow\right\rangle_l+\sin\chi\left|\downarrow\right\rangle_l\right)\\
&\left| \text{gs}\right\rangle_2=\prod_{l}\left(\cos\chi\left|\uparrow\right\rangle_l-\sin\chi\left|\downarrow\right\rangle_l\right),
\label{Eq:degeneracy}
\end{split}
\end{equation}
where $\cos^2(2\chi)=\frac{1-\gamma}{1+\gamma}$, and $\left|\uparrow\right\rangle_l$ ($\left|\downarrow\right\rangle_l$) are real spin states in the $\sigma_z$ basis at the site $l$.\\

\begin{figure}
\includegraphics{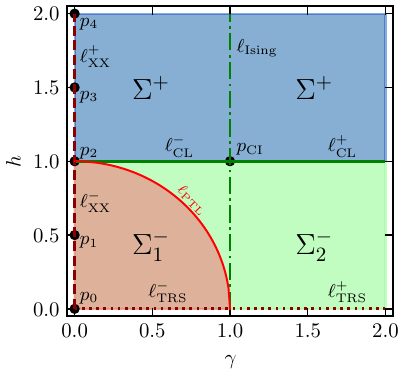}
\caption{Definition of the regions in the $\gamma-h$ phase space introduced in SEC.~\ref{SEC:definitions}.}
\label{Fig:Names}
\end{figure}

Due to the different properties of the XY chain, we subdivide the full $\gamma$-$h$ space to the following regions:
\begin{itemize}
    \item $\Sigma_1^-$ area: $\gamma^2+h^2<1$,
    \item $\Sigma_2^-$ area :$|h|<1$ and $\gamma^2+h^2>1$ and $\gamma\ne 0$,
    \item $\Sigma^+$ area: $|h|>1$,
    \item $\ell_{\text{PTL}}$: parity transition line $\gamma^2+h^2=1$,
    \item $\ell^-_{\text{CL}}$: critical lines, $h=\pm 1$ and $\gamma<1$,
    \item $\ell^+_{\text{CL}}$: critical lines, $h=\pm 1$ and $\gamma>1$,
    \item $\ell^{-}_{\text{TRS}}$: Time reversal symmetric line, $h=0$ and $0<\gamma<1$,
    \item $\ell^{+}_{\text{TRS}}$: Time reversal symmetric line, $h=0$ and $\gamma>1$,
    \item $\ell^-_{\text{XX}}$: $U(1)$-symmetric critical XX line: $\gamma=0$, $|h|\le 1$,
    \item $\ell^+_{\text{XX}}$: $U(1)$-symmetric non-critical XX line: $\gamma=0$ $|h|>1$,
    \item $\mathcal{P}_{\text{XX}}\equiv \left\lbrace  P_n=(0,\frac{n}{2})\ |\  n\in \mathbb{Z}\right\rbrace$,
    \item $\ell_{\text{Ising}}$: non-critical Ising line, $(\gamma=1, h\ne 1)$
    \item $p_{\text{CI}}$ point: Critical Ising point, ($\gamma=1$,$h=1$).
\end{itemize}
These regions are represented in fig.~\ref{Fig:Names}. 

\section{ANALYSIS OF NS AND R SECTORS}\label{3}

In this section, we discuss the energy spectrum and the ground state properties of the NS and R sectors in terms of $\gamma$ and $h$. 

\subsection{Energy Spectrum}
We start with the single-particle spectrum of the NS ($N_L=+1$) and R ($N_L=-1$) fermionic systems Eq.~\ref{Eq:SPEnergy}. The excitation energies $\epsilon(k)$ are shown in Fig. \ref{Fig:App:R-and-NS}, using of which one can calculate the total energy spectrum. The ground state energy for the fermionic sectors (Eqs.~\ref{Eq:HXYDiag} and~\ref{Eq:SPEnergy}) read
\begin{equation}
E_{N_L}^{(\text{gs})}=-\frac{1}{2}\sum_k\epsilon_{N_L}(k),
\label{Eq:FermionicEnergy}
\end{equation}
while the excited state energies are obtained using the relation
\begin{equation}
E_{N_L}=E_{N_L}^{(\text{gs})}+\sum_k n_k\epsilon_{N_L}(k),
\end{equation}
where $n_k$ is the number of fermions in the mode $k$. The first excited energy is therefore $E_{N_L}^{(\text{ES})}=E_{N_L}^{(\text{GS})}+\epsilon_{N_L}(k_{\text{min}})$, where $k_{\text{min}}$ is defined in Eq.~\ref{Eq:FES}.\\

Therefore, there are totally $2^L$ eigen energies for the NS fermionic sector (corresponding to the total number of ways of exciting single particles), and $2^L$ eigen energies for R fermionic sector. Given that there are $2^L$ eigen energies for the spin chain hamiltonian, one concludes that not all the (NS or R) fermionic eigen energies and the corresponding eigenstates belong to the energy spectrum of the real spin chain. Depending on the external parameters of the model, the ground state and the excited states of the spin chain belongs to the spectrum of the NS and R fermionic sectors.\\ 

One may draw a $\gamma$-$h$ phase diagram, which monitors the structure of the energy spectrum. For finite chains, only 5 possibilities were recovered listed as follows (the symbol $\longleftrightarrow$ exhibits the correspondence, and the subscript SC stands for ``spin chain"):  
\begin{itemize}
    \item \textbf{Case (1)} corresponds to the situation where the ground state of spin chain corresponds to the NS fermionic system, while the first excited state belongs to the R sector, i.e.
\begin{equation}
\begin{split}
\left\lbrace \begin{matrix}
|\text{gs}\rangle_{\text{SC}} \longleftrightarrow |\text{gs},N_L=+1\rangle\\
|\text{es}\rangle_{\text{SC}} \longleftrightarrow |\text{gs},N_L=-1\rangle 
\end{matrix}  \right. ,
\end{split}
\label{Eq:states}
\end{equation}
where the states in the left hand side belong to the spin chain, while the states in the right hand side are those for fermionic sectors. 
\item \textbf{Case (2)} shows the inverse situation:
\begin{equation}
\begin{split}
\left\lbrace \begin{matrix}
|\text{gs}\rangle_{\text{SC}} \leftrightarrow |\text{gs},N_L=-1\rangle\\
|\text{es}\rangle_{\text{SC}} \leftrightarrow |\text{gs},N_L=+1\rangle 
\end{matrix}  \right. .
\end{split}
\label{Eq:states}
\end{equation}
\item \textbf{Case (3)} corresponds to the situation where 
\begin{equation}
|\text{gs}\rangle_{\text{SC}} \leftrightarrow |\text{gs},N_L=+1\rangle ,
\end{equation} 
while none of the energy levels in the R sector does exist in the SC spectrum.
\item  \textbf{Case (4)} and \textbf{Case (5)} represent the situation where
\begin{equation}
|\text{gs}\rangle_{\text{SC}}\leftrightarrow |\text{gs},N_L=+1\rangle\leftrightarrow|\text{gs},N_L=-1\rangle
\end{equation}
(the ground state of the R and the NS sectors are the same, equal to the ground state energy of the spin chain). $|\text{gs}\rangle_{\text{SC}}$ is not degenerate for case (4) and is degenerate for case (5).
\end{itemize}

\begin{figure*}
\includegraphics[width=0.24\textwidth]{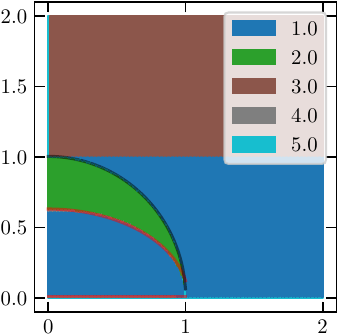}
\includegraphics[width=0.24\textwidth]{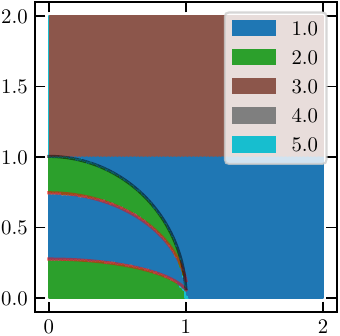}
\includegraphics[width=0.24\textwidth]{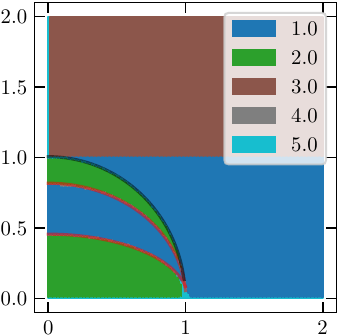}
\includegraphics[width=0.24\textwidth]{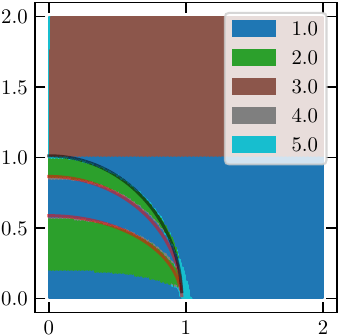}\\
\includegraphics[width=0.24\textwidth]{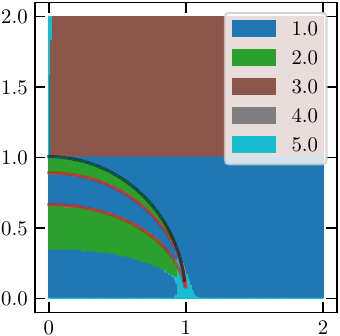}
\includegraphics[width=0.24\textwidth]{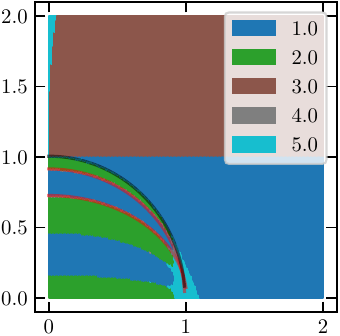}
\includegraphics[width=0.24\textwidth]{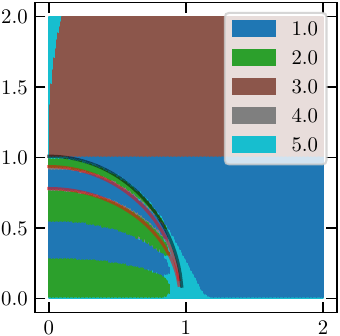}
\includegraphics[width=0.24\textwidth]{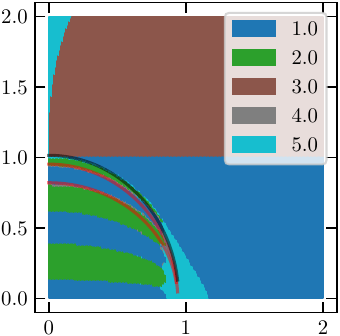}
\caption{A $\gamma-h$ (ground states and excited states) phase diagram in terms of the the system size in R and NS sectors. From top-left to bottom-right $L$ takes the values $L= 5,6,7,8,9,10,11$ and 12. Dark blue, green, brown, grey and light blue show respectively case (1), case (2), case (3), case (4) and case (5) as defined in Eq.~\ref{Eq:states} and the subsequent text.}
\label{Fig:PhaseDiagramEnergy}
\end{figure*}

An color map of the $\gamma$-$h$ phase space is represented in Fig.~\ref{Fig:PhaseDiagramEnergy} for $5\le L\le 12$, showing a strong $L$ dependence asexpected since $N_L$ depends on $L$. By ``phase'' we mean different cases described above. We see that different phases are sharply separated by some ``transition'' lines, or ``phase boundaries''. On $\ell_{\text{CL}}^{\pm}$ where the XY chain undergoes a magnetic transition between ferromagnetic and paramagnetic phases, the system transits between case (1) and case (3) in the fermionic sector, i.e. the ground state is in the NS sector for both. On $\ell_{\text{PTL}}$ (represented by bold-black lines in all figures) another transition is observed where the system goes from case 2 ($h<\sqrt{1-\gamma^2}$) to case 1 ($h>\sqrt{1-\gamma^2}$), while right on $\ell_{\text{PTL}}$ we have case 5, i.e. the spin chain ground state energy is degenerate equal to the ground state energy of NS and R sectors, which is expected from Eq.~\ref{Eq:degeneracy}. We also see other transition lines that describe successive transitions between states (1) and (2) showing that the role of $N_L=+1$ and $N_L=-1$ are interchanged successively. These transition lines are governed by the equation
\begin{equation}
\gamma^2+\left(\frac{h}{h_0(i,L)}\right)^2=1,
\label{Eq:circle2}
\end{equation}
where $h_0(i,L)$ is a parameter characterizing the $i$th transition line. On top of all the transition lines (except the upper most one corresponding to $\ell_{\text{PTL}}$) the status of the system is described by case 4. All these lines converge at the special point $(\gamma=1,h=0)$, where states (2) and (3) coexist. The number of transition lines, $M(L)$ depends on $L$ as given by
\begin{equation}
M(L)=\left[\frac{L}{2}\right],
\end{equation}
where $[x]$ represents the integer part of $x$.  The parity transition line $\ell_{\text{PTL}}$ is especial in the sense that it intersects with the ferromagnetic transition line $h=h_c$ at $\gamma=0$, i.e. showing $i=M(L)$ for the $\ell_{\text{PTL}}$ line:
\begin{equation}
h_0(M(L),L)=h_c.
\end{equation} 
Our observations reveal that, while the thermodynamic limit is an important leading limit, some physics of finite quantum systems are lost in this limit. Extrapolating our results to $L\to\infty$ reveals that the ground state of the XY quantum chain is erratic and is more similar to random in $\Sigma_1^-$, while it is NS in $\Sigma_2^-\cup\Sigma^+$. Note that the expansion of regions with status case (5) for $h>h_c$ as $L$ increases is due to numerical restrictions.

\subsection{$L$-dependence of the energy scales}\label{SEC:EnergySpec}
Although the quantities associated with NS and R fermions are expected to converge in the thermodynamic limit, the key question is how this convergence occurs in finite systems. Specifically, we focus on the difference between the NS and R ground state energies and its dependence on $L$
\begin{equation}
\begin{split}
\delta E^{(\text{GS})} & \equiv E^{(\text{GS})}_{N_L=+1}-E^{(\text{GS})}_{N_L=-1}.
\end{split}
\label{Eq:delta_E}
\end{equation}

\begin{figure}
\centering
\includegraphics[scale=1.0]{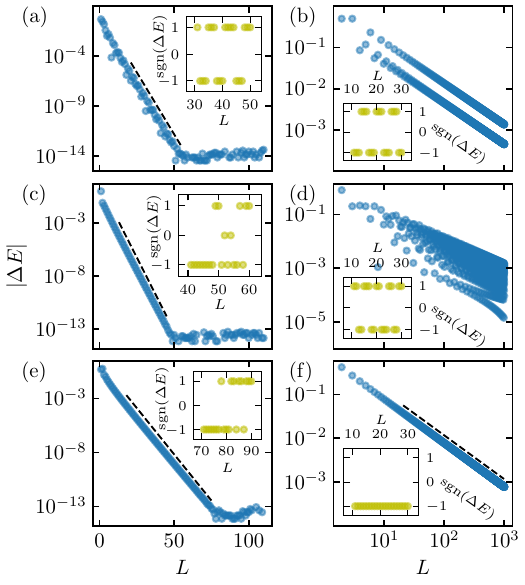}
\caption{The absolute value (main panel) and the sign (inset panel) of the difference between the ground state energies of the NS and the R sectors for the ordered pairs of $(\gamma,h):$ (a) $(0.5,0.5)$, (b) $(0,0.5)$, (c) $(0.5,1.5)$, (d) $(0,0.2)$, (e) $(1.5,0.5)$, and (f) $(1,1)$. Note that the scale of the main panels in (a), (c) and (e) is  log-normal, while that of the (b), (d) and (f) is log-log.}
\label{Fig:delta-E}
\end{figure}
Figure~\ref{Fig:delta-E} presents the absolute value and sign of the energy difference between the R and NS sectors. The first column uses a semi-log scale, where a linear trend indicates exponential decay, while the second column displays the data on a log-log scale, where a linear trend suggests power-law behavior. The insets illustrate the sign of $\delta E^{(\text{GS})}$, revealing an alternating pattern in which the ground state oscillates between the NS and R sectors. This oscillatory behavior suggests a competition between the NS and R ground states, particularly in smaller chains, which diminishes as $L$ increases. Notably, in non-critical cases ($h\ne 1$ and $\gamma\ne 0$), the decay follows an exponential trend
\begin{equation}
\left| \delta E^{(\text{GS})}(\gamma\ne 0,h\ne h_c)\right| 	 \sim \exp\left[-\beta^{(E)}_{\gamma,h} L\right],
\label{Eq:exponential}
\end{equation}
where $\beta^{(E)}_{\gamma,h}$ is a $(\gamma,h)$-dependent decay exponent. For the critical case (ferromagnetic transition point $h=h_c$) however the decay is found to be power-law:
\begin{equation}
\left| \delta E^{(\text{GS})}(\gamma\ne 0,h=h_c)\right| 	 \sim L^{-\alpha_{\gamma}^{(E)}},
\label{Eq:power-Law}
\end{equation}
where $\alpha_{\gamma}^{(E)}$ is another scaling exponent which depends on $\gamma$. For $\gamma=0$ the energy difference is erratic.\\
\begin{figure*}
\centering
\includegraphics[width=0.32\linewidth]
{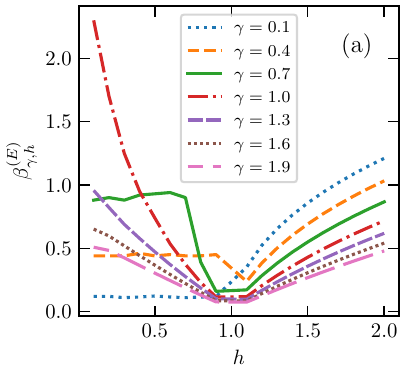}
\includegraphics[width=0.32\linewidth]
{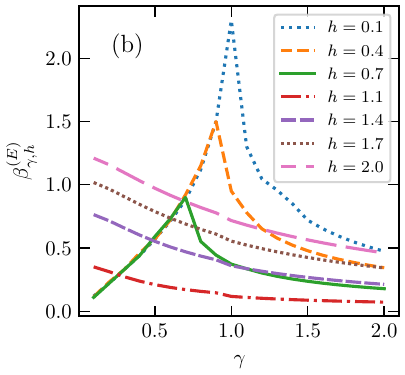}
\includegraphics[width=0.32\linewidth]
{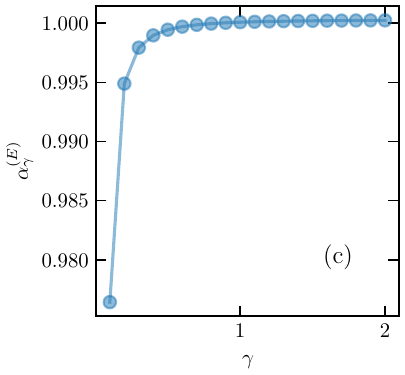}
\caption{(a) and (b) show the variation of the exponential exponent, $\beta^{(E)}_{\gamma,h}$, defined in Eq.~\ref{Eq:exponential} for different values of $\gamma$ and $h$ while (c) represents the dependence of $\alpha^{(E)}_{\gamma}$ on $\gamma$, defined in Eq.~\ref{Eq:power-Law}, for $h=h_c=1$.}
\label{fig:dE-alpha}
\end{figure*}

The dependence of the ``decay rate" $\beta_{\gamma,h}^{(E)}$ on $\gamma$ and $h$ is illustrated in Fig.~\ref{fig:dE-alpha}, excluding $h=h_c$ due to its power-law scaling with $L$. For fixed $\gamma$, $\beta_{\gamma,h}^{(E)}$ increases monotonically with $h$ for $h>h_c$, whereas for $h<h_c$ its dependence on $h$ is non-monotonic, reaching a minimum near $h=h_c$. This indicates that the decay slows down around the critical point. Figure~\ref{fig:dE-alpha}b displays $\beta_{\gamma,h}^{(E)}$ as a function of $\gamma$ for different values of $h$. Notably, $\beta_{\gamma,h}^{(E)}$ increases in $\Sigma^-_1$ and decreases in $\Sigma^-_2\cup\Sigma^+$. Notice the change of behavior for $h<1$ and $h>1$, the former characterized by sharp cusp and an absolute maximum along $\ell_{\text{PTL}}$, where the ground state parity shifts. This suggests that the NS and R ground state energies converge most rapidly on $\ell_{\text{PLT}}$ where the decay exponent is at its largest. \\

\begin{figure}[t]
\centering
\includegraphics[width=6cm, clip]{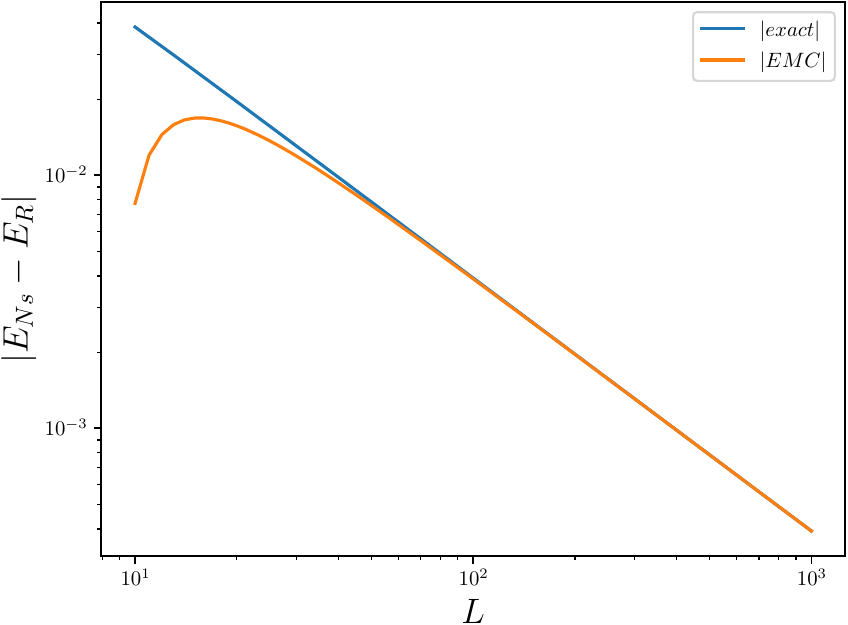}
\caption{Log-Log plot of the energy difference behavior for $h = h_c$ and $\gamma = 0.5$. The blue curve represents the EMC approximation, while the orange curve corresponds to the exact function.}
\label{Fig:EMC}
\end{figure}

Figure~\ref{fig:dE-alpha}c illustrates the dependence of $\alpha_{\gamma}^{(E)}$ on $\gamma$ along $\ell_{\text{CL}}$ ($h=h_c$), showing that it remains close to unity, with deviations appearing for $\gamma$ values near zero. This implies that for $\gamma\ne 0$ $\alpha_{\gamma}^{(E)}=1$. To understand this behavior, we note that in the critical region, the summation in Eq.~\ref{Eq:delta_E} closely approximates its integral form, which becomes exact in the thermodynamic limit $L\to \infty$, $\text{d}k\to \frac{L}{2\pi}\text{d}\phi$: 
\begin{equation}
E^{(\text{GS})}_{L\to\infty}=-\frac{L}{4\pi}\int_0^{2\pi} \text{d}\phi \sqrt{(h-\cos\phi)^2+\gamma^2\sin^2\phi}.
\end{equation}
In this case one expects that the Euler-Maclaurin formula~\cite{apostol1999elementary} applies for large but finite $L$ values, see appendix \ref{The EMC expansion theorem} for details. 
Writing $\delta E^{(\text{GS})}$ as
\begin{equation}
\delta E^{(\text{GS})}=(E^{(\text{GS})}_{N=+1}-E^{(\text{GS})}_{L\to\infty})-(E^{(\text{GS})}_{N=-1}-E^{(\text{GS})}_{L\to\infty}),
\end{equation}
one uses the Euler-Maclaurin formula Eq.~\ref{Eq:AppB}, which to the leading order reads
\begin{equation}
\begin{split}
E^{(\text{GS})}_{N}-E^{(\text{GS})}_{L\to\infty}=&-\frac{1}{2}\left[\frac{\epsilon_N(1)+\epsilon_N(L)}{2}+\frac{1}{6} \frac{\epsilon'_N(1)+\epsilon'_N(L)}{2!}\right]\\
&+...,
\end{split}
\end{equation}
where $\epsilon'\equiv \frac{\text{d}}{\text{d}k}\epsilon(k)$. This gives ($\gamma\ne 0$) 
\begin{equation}
\delta E^{(\text{GS})}_{L\to\infty}=\frac{\pi \gamma}{12}\left(\frac{1}{L}\right) - \frac{61 \pi^3 (4\gamma^2 - 3)}{720 \gamma} \left(\frac{1}{L^3}\right)+O(L^{-5}),
\label{Eq:EnergyDifference}
\end{equation}
which is demonstrated in Fig.~\ref{Fig:EMC} for $h=h_c$ and $\gamma=0.5$, where excellent agreement with exact numerical data is observed. The discrepancy at small $L$ values arises from sub-leading contributions at higher orders of $L$.

From this analysis, we conclude that the convergence of the NS and R ground state energies is slow (following a power law) in critical cases, whereas for non-critical cases, the convergence is exponential. Notably, the fastest (exponential) convergence occurs along the $\ell_{\text{PTL}}$ line.

\section{GEOMETRY OF XY CHAIN}\label{4}

In this section, we provide a geometrical interpretation of quantum states in spin chains using Berry phases. The quantum geometric properties of spin chains are expected to be affected by the system's finite size, and the behavior of R and NS sectors in finite systems are expected to be different. Specifically, the values of the quantum metrics vary between these sectors, especially near critical points, indicating that boundary conditions have a significant impact on the geometric properties of the system. The sensitivity of geometric measures to boundary conditions underscores the importance of considering these effects in practical implementations, such as quantum computing and simulation platforms~\cite{dutta2015quantum}.

\subsection{Berry Phase and Criticality}

In quantum systems, the Berry phase reveals essential information about the critical behavior without relying on an order parameter or symmetry-breaking analysis~\cite{Berry1984Quantal, zhu2006scaling}. Consider a quantum system described using the Hamiltonian $H(\boldsymbol{\lambda})$ that depends on a set of parameters $\boldsymbol{\lambda} = (\lambda_1, \lambda_2, ...,\lambda_d)$ ($d$ representing the dimension of the slow-parameter space), with an eigenstate $\left| \psi(\lambda)\right\rangle$ evolve adiabatically as the parameters vary slowly. In our paper $\lambda_1\equiv \gamma$, and $\lambda_2\equiv h$. The resulting Berry phase acquired by the system when the parameters trace a closed loop $C$ is given by:
\begin{equation}
    \phi_{\psi}^{\mathcal{B}}(C) = \oint_C A_\mu^{\mathcal{B}} d\lambda_\mu,
\end{equation}
where $A_\mu^{\mathcal{B}} = i \langle \psi(\lambda) | \frac{\partial}{\partial \lambda_\mu} | \psi(\lambda) \rangle$ represents the Berry connection. The Berry phase  becomes particularly significant near critical points, where singularities arise, and induces singularities in the other higher order geometrical quantities~\cite{zanardi2007information}. \\

The quantum geometric tensor is a fundamental quantity in quantum information geometry that captures the response of a quantum system's ground state to changes in external parameters~\cite{provost1980riemannian,kolodrubetz2017geometry}. It is defined as ($\partial_\mu\equiv \frac{\partial}{\partial \lambda_\mu}$ and $| \psi \rangle \equiv | \psi(\lambda) \rangle$):
\begin{equation}
    Q_{\mu\nu} = \langle \partial_{\mu}\psi | \partial_{\nu}\psi \rangle - \langle \partial_{\mu}\psi | \psi \rangle \langle \psi | \partial_{\nu}\psi \rangle,
\end{equation}
where $|\psi\rangle$ is considered to be the ground state in this study. The quantum geometric tensor is generally a complex-valued tensor:
\begin{equation}
    Q_{\mu\nu} = g_{\mu\nu} - i \Omega_{\mu\nu}.
\end{equation}
where $g_{\mu\nu}$ defines a Riemannian metric on the quantum state space, often called Fubini–Study metric and $\Omega_{\mu\nu}$ is the Berry curvature, which is associated with geometric phases acquired during adiabatic evolutions~\cite{berry1989quantum,campos2007quantum}:
\begin{equation}
\Omega_{\mu\nu}\equiv \partial_\mu A^{\mathcal{B}}_\nu-\partial_\nu A^{\mathcal{B}}_\mu.
\end{equation}
$g_{\mu\nu}$ is also related to the quantum Fisher information matrix $\mathcal{F}_Q$ for the pure states, which governs the quantum Cram\'er–Rao bound~\cite{paris2009quantum}:
\begin{equation}
\text{Cov}[\hat{\theta}]\ge \frac{1}{M} \mathcal{F}_Q^{-1},
\end{equation}
where 
$M$ is the number of measurements, and $\text{Cov}[\hat{\theta}]$ is the covariance matrix of any unbiased estimator. The inequality means the covariance matrix of any unbiased estimator is bounded below by the inverse of the quantum Fisher information matrix. $g_{\mu\nu}$ is also related to fidelity susceptibility tensor according to~\cite{zanardi2007information}
\begin{equation}
\begin{split}
F(\boldsymbol{\lambda},\boldsymbol{\lambda}+\text{d}\boldsymbol{\lambda}) & \equiv \langle \psi(\boldsymbol{\lambda}) | \psi(\boldsymbol{\lambda}+\text{d}\boldsymbol{\lambda}) \rangle\\
& \approx 1-\frac{1}{2}\sum_{\mu\nu}g_{\mu\nu}\text{d}\lambda_\mu\text{d}\lambda_\nu,
\end{split}
\end{equation}
so that the corresponding fidelity susceptibility in the $\textbf{v}=(v_1,v_2)$ direction is defined as
\begin{equation}
\chi_{F}^{(\textbf{v})}=\sum_{\mu\nu}g_{\mu\nu}v_\mu v_\nu.
\end{equation}
We observe that the Fubini-Study metric encodes the fidelity susceptibility, capturing the system's sensitivity to parameter changes—that is, how rapidly the ground state shifts direction in Hilbert space as the system parameters are varied. In quantum phase transitions, this quantity often peaks or diverges near quantum critical points, i.e. plays a similar role as heat or magnetic susceptibility in classical system which are used to detect phase transitions~\cite{yu2009fidelity}.\\

The Fubini-Study metric and quantum entanglement are deeply interconnected through the geometry of quantum state space. While a universal, explicit formula linking the Fubini-Study metric directly to entanglement entropy doesn't exist, several studies have explored their relationship, especially in multipartite systems and near critical points. In multipartite systems, changes in entanglement due to variations in parameters 
can lead to significant changes in the Fubini-Study metric, indicating a sensitivity to entanglement properties~\cite{zanardi2007information,you2007fidelity,campos2007quantum,rajabpour2017multipartite}. For example the entanglement distance was found to depend directly to Fubini-Study metric~\cite{vesperini2024unveiling}.\\

The Ricci scalar $\mathcal{R}$, derived from the metric tensor, provides a global measure of the manifold's curvature and reveals crucial information about the phase structure of the system~\cite{Hernando2007Geo}. For the two-dimensional space $(\gamma,h)$ in this paper $\mathcal{R}$ is given by the following equation:
\begin{equation} 
\begin{split}
\mathcal{R} =-\frac{1}{2 }&\begin{vmatrix}
   g_{hh} & g_{h\gamma} \\
   g_{h\gamma} & g_{\gamma\gamma}
\end{vmatrix}^{-2}\times\begin{vmatrix}
   g_{hh} & g_{h\gamma} & g_{\gamma\gamma}\\
   \frac{\partial g_{hh}}{\partial h} & \frac{\partial g_{h\gamma}}{\partial h} & \frac{\partial g_{\gamma\gamma}}{\partial h}\\
   \frac{\partial g_{hh}}{\partial \gamma} & \frac{\partial g_{h\gamma}}{\partial \gamma} & \frac{\partial g_{\gamma\gamma}}{\partial \gamma}
\end{vmatrix}\\
=\frac{1}{\sqrt{g}} &\left( \frac{\partial}{\partial h} \frac{\sqrt{g}}{g_{\gamma\gamma}} \frac{\partial g_{hh}}{\partial \gamma} + \frac{\partial}{\partial \gamma} \frac{\sqrt{g}}{g_{hh}} \frac{\partial g_{\gamma\gamma}}{\partial h} \right.\\
&\left. -\frac{\sqrt{g}}{2 g_{hh} g_{\gamma\gamma}} \frac{\partial g_{hh}}{\partial \gamma} \frac{\partial g_{\gamma\gamma}}{\partial h} \right),
\end{split}
\end{equation}
where $ g = \det(g_{\mu\nu})$, and $|...|$ symbolizes the determinant. An analytical evaluation of this expression suggests that $\mathcal{R} $ exhibits divergent behavior at certain critical points, indicating its potential as a marker of quantum phase transitions~\cite{zanardi2007information,johnston2003information}. Additionally, the sign of $\mathcal{R} $ provides valuable insights into the nature of the interactions within the system: positive curvature indicates dominant repulsive interactions, while negative curvature corresponds to attractive interactions~\cite{janyszek1989riemannian, ruppeiner1995riemannian}.

In finite-size systems, the behavior of $\mathcal{R} $ is distinct in the NS and R sectors. In certain parameter regimes, $\mathcal{R} $ can be positive in one sector and negative in the other, reflecting different interaction characteristics and stability profiles. Divergences and sign changes in $\mathcal{R} $ also serve as clear indicators of phase transitions, with finite-size scaling analyses providing estimates for critical exponents related to these transitions~\cite{campos2007quantum}.

The differences in geometric quantities between the NS and R sectors have important physical implications. The divergences in $\mathcal{R} $ identify the points that are candidates for phase transitions. Also, understanding the geometric structure of the system can be valuable for optimizing quantum protocols that rely on adiabatic evolution, and this knowledge can also inform strategies for error correction in quantum information applications~\cite{pastawski2015holographic, kitaev2003fault}, see Appendix \ref{CURVATURE OF THE PARAMETER SPACE}.

\subsection{The XY Model Geometry in the Thermodynamic Limit}

In the thermodynamic limit ($L\to \infty$), the geometric properties such as the Berry phase exhibitS the following asymptotic behavior~\cite{carollo2020geometry}:
\begin{equation}
    \phi^{\mathcal{B}}_{\text{XY}}(L\to\infty) =  \begin{cases}
       -\pi + \frac{\pi h \gamma}{\sqrt{(1 - \gamma^2)(1 - \gamma^2 - h^2)}}&\text{for } (\gamma,h)\in \Sigma^-_1\\
       0 &\text{otherwise}
    \end{cases},
\end{equation}
while the divergence of $\mathcal{R}$ becomes more pronounced, allowing for a precise identification of critical behavior. In this limit~\cite{zanardi2007information}
\begin{widetext}
\begin{equation}
 \mathcal{R} \to \begin{cases}
-\frac{4}{L} \left(1+| \gamma |\right) /| \gamma | &\text{for } (\gamma,h)\in \Sigma_1^- \ \text{or}\  \Sigma_2^-  \\
\frac{4}{L} \left( |h|+ \sqrt{h^2 + \gamma^2 - 1} \right)/\sqrt{h^2 + \gamma^2 - 1} &\text{for } (\gamma,h)\in \Sigma_1^+.
\end{cases}
\label{Eq:RThermodynamicLimit}
\end{equation}
\end{widetext}
Note that the curvature diverges on the segment $|h|<1$, $\gamma=0$ ($\ell_{\text{XX}}^-$) and is discontinuous on the lines $\ell_{\text{CL}}$. It confirms the known fact that $\lim_{|h|\to 1^+}\mathcal{R}=-\lim_{|h|\to1^-}\mathcal{R}$. These divergences, particularly along the segment $\ell_{\text{XX}}^-$, are hallmarks of criticality in the XY model, offering a deep connection between geometric properties and the physical behavior of the system near critical points.

\section{GEOMETRY OF FINITE XY CHAIN}\label{SEC:Geometry}

In this section, we investigate the geometrical properties of the NS and R sectors of finite XY spin chains, focusing on the behavior of the quantum geometric tensor $Q_{\mu\nu}$ and its associated curvature. The details of the calculation of the components of $Q_{\mu\nu}$ are presented in the appendix~\ref{Metric Elements}. The three elements are found to be as follows: 
\begin{equation}
\begin{aligned}
Q_{hh} =&  \frac{1}{4}\sum_{k=1}^L \left(\frac{\partial\theta^{(N_L)}_{k}}{\partial h}\right)^{2}\\
=&\frac{1}{4}\sum_{k=1}^L \left(\frac{-\gamma \sin\phi^{(N_L)}_{k}}{(h-\cos\phi^{(N_L)}_{k})^2 + \gamma^2 \sin^2\phi^{(N_L)}_{k}}\right)^2,\\
Q_{\gamma \gamma} = & \frac{1}{4}\sum_{k}  \left(\frac{\partial\theta^{(N_L)}_{k}}{\partial \gamma}\right)^{2}
 \\&= \frac{1}{4}\sum_{k=1}^L \left(\frac{\sin\phi^{(N_L)}_{k}(h-\cos\phi^{(N_L)}_{k})}{(h-\cos\phi^{(N_L)}_{k})^2 + \gamma^2 \sin^2\phi^{(N_L)}_{k}}\right)^2,\\
 Q_{h \gamma} = & \sum_{k} \frac{1}{4} \left(\frac{\partial\theta^{(N_L)}_{k}}{\partial h}\right)\left(\frac{\partial\theta^{(N_L)}_{k}}{\partial \gamma}\right)\\
=& \frac{1}{4}\sum_{k=1}^L \left(\frac{-\gamma\sin^2\phi^{(N_L)}_{k}(h-\cos\phi^{(N_L)}_{k})}{[(h-\cos\phi^{(N_L)}_{k})^2 + \gamma^2 \sin^2\phi^{(N_L)}_{k}]^2}\right) .
\end{aligned}
\end{equation} 
By evaluating these summations, one finds singular and non-singular behaviors of the system~\cite{zanardi2007information}.

\begin{figure*}
    \centering
    \includegraphics{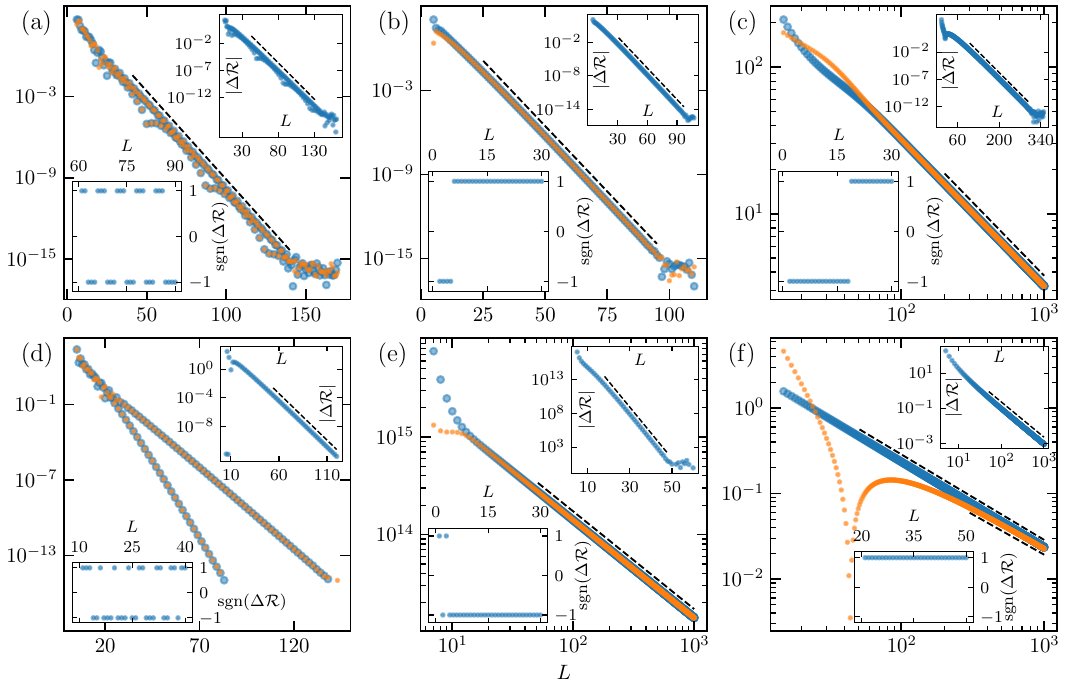}
        \caption{Absolute values of curvatures, $|\mathcal{R}_R|$ (blue disks) and $|\mathcal{R}_{NS}|$ (orange points), in terms of $L$ for ordered pairs of $(\gamma, h)$: (a) (0.3, 0.5), (b) (1.3, 0.5), (c) (0.7, 1.1), (d) (0.3, 0), (e) (0, 1.5), and (f) (1.2, 1). While the upper insets in each of these panels show the variation of $|\Delta \mathcal{R}|=|\mathcal{R}_{NS}-\mathcal{R}_R|$ with $L$ for the corresponding $(\gamma, h)$ values, the lower insets reveal values of $L$ for which the sign of $|\Delta \mathcal{R}|$ changes.}
\label{Fig:curvature_combined}
\end{figure*}
\begin{figure*}[t]
\centering
{\includegraphics[width=0.325\linewidth]
{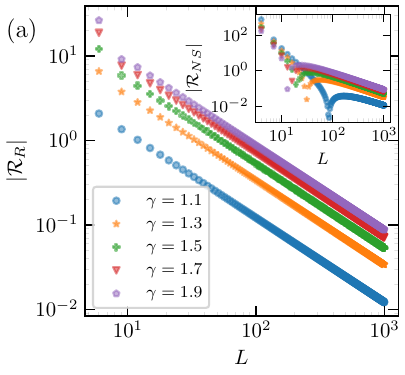}}
{\includegraphics[width=0.325\linewidth]
{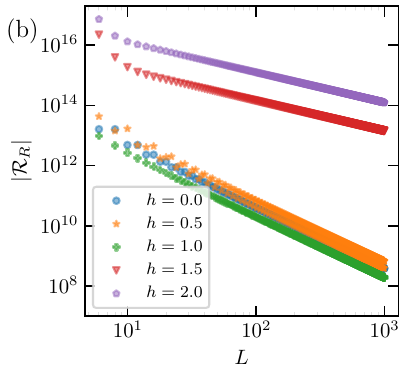}}
{\includegraphics[width=0.325\linewidth]
{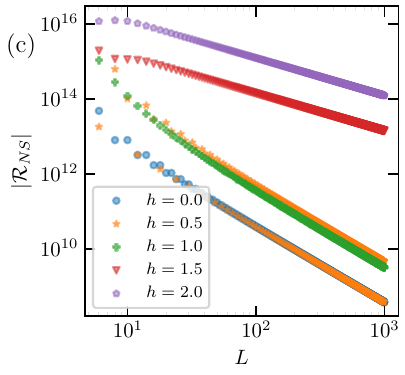}}
\caption{(a) Variation of $|\mathcal{R}_R|$ (main figure) and $|\mathcal{R}_{NS}|$ (inset) as a function of $L$ when $h=h_c=1$ and $1<\gamma<2$. Corresponding figures for $\gamma=0$ of (b) the Ramond sector and (c) Neveu-Schwarz sector for different $h$ values.}
\label{fig:h1}
\end{figure*}
Figure~\ref{Fig:curvature_combined} shows the curvature in terms of $L$ for six points in the phase space: $(\gamma,h)= (0.3, 0.5)$ which is a representative point of $\Sigma_1^-$, $ (1.3, 0.5)$ as a representative of $\Sigma_2^-$, $(0.7, 1.1)$ as a representative of $\Sigma^+$, $(0.3, 0)$ as a representative of $\ell_{\text{TRS}}$, $(0, 1.5)$ as a representative of $\ell_{\text{XX}}^+$, and  $(1.2, 1)$ as a representative of $\ell_{\text{CL}}$. In each figure, the main part represents $|\mathcal{R}_{\text{R}}|$ and $|\mathcal{R}_{\text{NS}}|$, the upper inset shows $|\Delta \mathcal{R}|\equiv |\mathcal{R}_{\text{R}}-\mathcal{R}_{\text{NS}}|$, and the lower inset shows the sign of $\Delta \mathcal{R}$. For the region $\Sigma_1^-\cup\Sigma_2^-\cup \ell_{{\text{PTL}}}$ the decay of the absolute value of curvature for both R and NS sectors ($|\mathcal{R}_{\text{R}}|$ and $|\mathcal{R}_{\text{NS}}|$ respectively), as well as $\Delta \mathcal{R}\equiv \mathcal{R}_{\text{R}}-\mathcal{R}_{\text{NS}}$ is exponential (Fig.~\ref{Fig:curvature_combined}a and~\ref{Fig:curvature_combined}b). The sign of $\Delta\mathcal{R}$ regularly changes in $\Sigma_1^-$, while it is positive for large enough systems in $\Sigma_2^-$. We observe that $|\mathcal{R}_{\text{R}}|$ and $|\mathcal{R}_{\text{R}}|$ decay in a power-law fashion, while $|\Delta\mathcal{R}|$ decays exponentially in the $\Sigma^+$ region, Fig.~\ref{Fig:curvature_combined}c. The behavior of the curvature is quite different on the $\ell_{\text{TRS}}$ line: Figure~\ref{Fig:curvature_combined}d shows that $|\mathcal{R}_{\text{R}}|$ and $|\mathcal{R}_{\text{NS}}|$ oscillate between two different exponentially decaying functions, while $\Delta\mathcal{R}$ decays with a single exponential function. On the $\ell_{\text{CL}}^-\cup\ell_{\text{CL}}^+$ line, all the curvatures and $\Delta\mathcal{R}$ decay in a power-law fashion. The only exception is $\mathcal{R}_{\text{NS}}$ on the $\ell_{\text{CL}}^+$ which changes sign at some $L$ (see Fig.~\ref{Fig:curvature_combined}f, featuring sharp change for the absolute value). This special $L$ moves to left as $\gamma$ increases (Fig.~\ref{fig:h1}a). On the $\ell_{\text{XX}}^-\cup\ell_{\text{XX}}^+$ $\mathcal{R}_{\text{R},\text{NS}}$ typically diverge, leading to an irregular $L$ dependence. The only exceptions on $\ell_{\text{XX}}^{\pm}$ are
\begin{equation}
(\gamma,h)\in \mathcal{P}_{\text{XX}}\equiv \left\lbrace  P_n=(0,\frac{n}{2})\ |\  n\in \mathbb{Z}\right\rbrace, \end{equation}
where the dependence is power-law decay. Figures~\ref{fig:h1}b and~\ref{fig:h1}c show the results for $n=0,1,2,3,4$.\\ 

We summarize these results in the following formulas:
\begin{equation}
\mathcal{R}_{\text{X}}\propto \begin{cases}
\exp \left[-\beta_{\text{X}}^{\mathcal{R}} L\right] &\text{for } (\gamma,h)\in \Sigma_{1}^-\cup\Sigma_{2}^-\cup\ell_{\text{PTL}},  \\
\infty & \text{for } (\gamma,h)\in \ell_{\text{XX}}^-\cup\ell_{\text{XX}}^+/\mathcal{P}_{\text{XX}},\\
L^{-\alpha_{\text{X}}^{\mathcal{R}}}  &\text{for } (\gamma,h)\in \Sigma^+\cup\ \ell_{\text{CL}}^-\cup\ \ell_{\text{CL}}^+\cup \mathcal{P}_{\text{XX}}.
\end{cases}
\label{Eq:Rx1}
\end{equation}
where X=NS and R. The behavior of $\mathcal{R}$ on the $\ell_{\text{TRS}}^{\pm}$ is quite different. In this case we have two branches where the curvature transits between them according to the $L$ value. More precisely
\begin{equation}
\mathcal{R}_X(\ell_{\text{TRS}}^-)\propto \begin{cases}
(-1)^{\frac{L}{2}+1}\exp \left[-\beta_X^{(\text{U})} L\right] & \text{mod}(L,4)=0, 2, \\
\exp \left[-\beta_X^{(\text{L})} L\right] &\text{mod}(L,4)=1, 3,
\end{cases}
\label{Eq:ROscillating1}
\end{equation} 
where $\beta_X^{(\text{U})}$ and $\beta_X^{(\text{L})}$ are two new exponents defined on $\ell_{\text{TRS}}^{\pm}$ lines, ``U'' standing for the upper branch and ``L'' standing for the lower branch, with the condition $\beta_X^{(\text{U})}<\beta_X^{(\text{L})}$. We call this function as ``bi-exponential". For $\ell_{\text{TRS}}^-$ we have $\beta_{\text{NS}}^{(\text{L,U})}=\beta_{\text{R}}^{(\text{L,U})}$, the exponents of the NS and R sectors are the same. For $\ell_{\text{TRS}}^+$ the bi-exponential behavior is different:
\begin{equation}
\mathcal{R}_X(\ell_{\text{TRS}}^+)\propto \begin{cases}
(-1)^{\frac{N_L-1}{2}}\exp \left[-\beta_X^{(\text{U})} L\right] & \text{mod}(L,2)=0, \\
\exp \left[-\beta_X^{(\text{L})} L\right] &\text{mod}(L,2)=1.
\end{cases}
\label{Eq:ROscillating2}
\end{equation} 
Note that $N_L=+1$ $(-1)$ for X=NS (R) sector. The difference of the Ricci scalers behaves like
\begin{equation}
\Delta\mathcal{R}\propto \begin{cases}
\exp \left[-\beta_{\Delta\mathcal{R}} L\right] &\text{for } (\gamma,h)\in \Sigma_1^-\cup\Sigma_2^-\cup\ell_{\text{PTL}}\cup\Sigma^+  \\
\text{Indetermined} & \text{for } (\gamma,h)\in \ell_{\text{XX}}^-\cup\ell_{\text{XX}}^+/ \mathcal{P}_{\text{XX}},\\
L^{-\alpha_{\Delta\mathcal{R}}}  &\text{for } (\gamma,h)\in \ell_{\text{CL}}^-\cup\ell_{\text{CL}}^+\cup \mathcal{P}_{\text{XX}}.
\end{cases}
\label{Eq:Rx2}
\end{equation} 

\begin{figure*}[t]
\centering
\includegraphics[width=0.32\linewidth]
{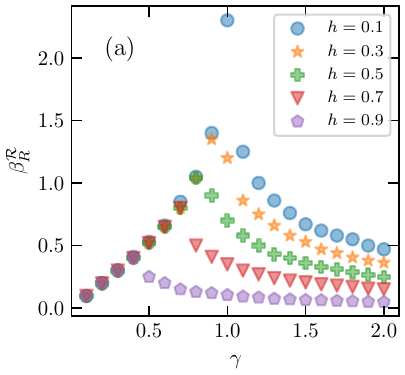}
\includegraphics[width=0.32\linewidth]
{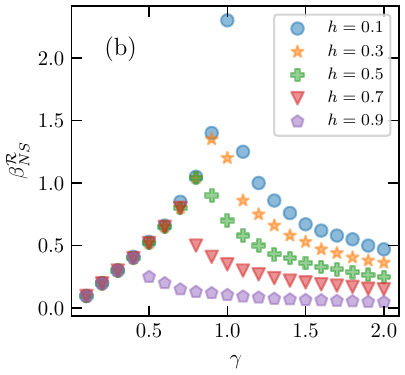}
\includegraphics[width=0.32\linewidth]
{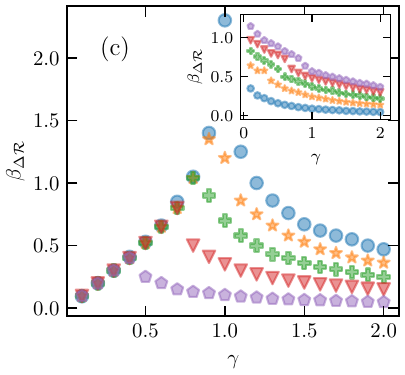}
\includegraphics[width=0.32\linewidth]
{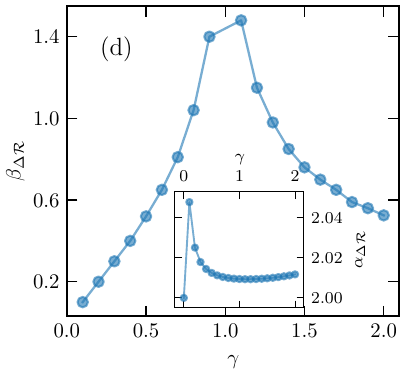}
\includegraphics[width=0.32\linewidth]
{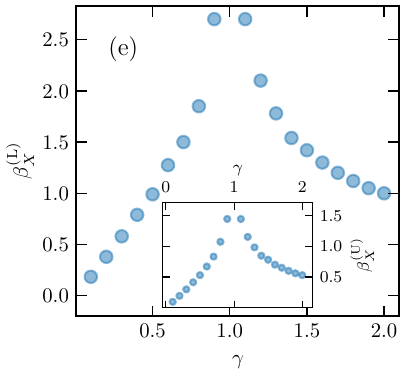}
{\includegraphics[width=0.32\linewidth]
{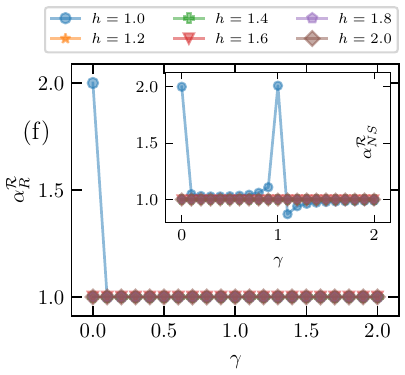}}
\caption{$\gamma$-dependent variation of the exponents: (a) $\beta^{\mathcal{R}}_R$, in the region $\Sigma_1^-\cup \Sigma_2^-$, (b) $\beta^{\mathcal{R}}_{NS}$, in the region $\Sigma_1^-\cup \Sigma_2^-$, (c) $\beta_{\Delta\mathcal{R}}$, in the regions $\Sigma_1^-\cup \Sigma_2^-$ (main panel) and $\Sigma^+$ (inset panel where higher $h$ values show higher exponents), (d) $\beta_{\Delta\mathcal{R}}$ when $h=0$ (main panel) and $\alpha_{\Delta\mathcal{R}}$ when $h=h_c=1$ (inset panel), (e) $\beta^{(L)}_X$ (main panel) and $\beta^{(U)}_X$ (inset panel) both for the case of $h=0$, and (f) $\alpha^{\mathcal{R}}_R$ (main panel) and $\alpha^{\mathcal{R}}_{NS}$ (inset panel) both in the region of $\Sigma^+ \cup \ell^-_{\mathrm{CL}} \cup \ell^+_{\mathrm{CL}}$. For definitions refer to the text.}
\label{fig:exponential}
\end{figure*}

\begin{figure*}[t]
\centering
\includegraphics[width=0.325\linewidth]
{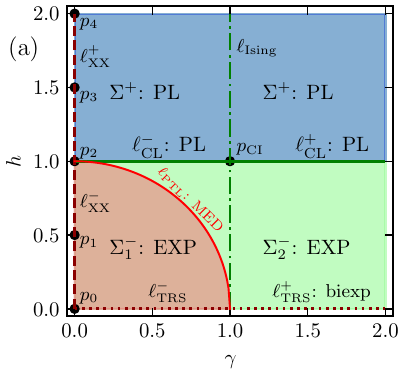}
\includegraphics[width=0.325\linewidth]
{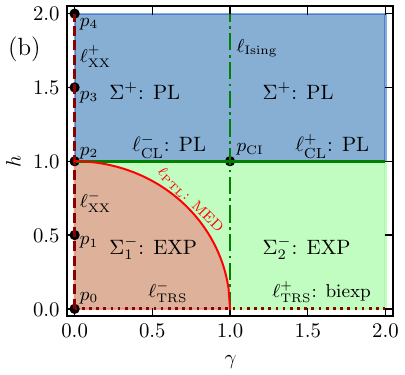}
\includegraphics[width=0.325\linewidth]
{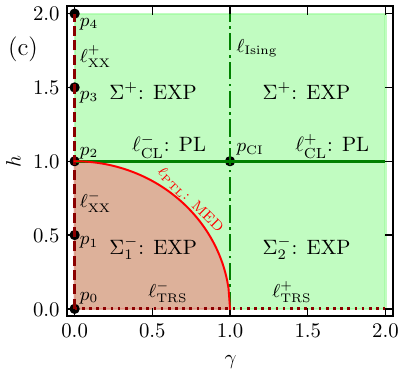}
\caption{Here, we show the important regions, lines, and points identified by analyzing the dependence form of (absolute value of) curvatures on $L$. PL refers to power-low, EXP refers to exponential decay, biEXP refers for bi-exponential decay introduced in Eqs.~\ref{Eq:ROscillating1} and~\ref{Eq:ROscillating2}, and MED refers to maximal exponential decay, stating that on the $\ell_{\text{PTL}}$ the exponent of the exponential decal is maximal. (a) shows the the Ramond sector, (b) the Neveu-Schwarz sector, and (c) their difference. For definitions refer to SEC.~\ref{SEC:definitions}.}
\label{fig:schematic}
\end{figure*}

It is worth-mentioning that these relations generalizes, and even contradicts in some cases with Eq.~\ref{Eq:RThermodynamicLimit} (which corresponds to $\alpha^{\mathcal{R}}=1$), stating that the curvature shows a rich behavior with the system size $L$ in terms of $(\gamma,h)$. The exponents $\alpha_{\text{R}}^{\mathcal{R}}$, $\alpha_{\text{NS}}^{\mathcal{R}}$, $\beta_{\text{R}}^{\mathcal{R}}$, $\beta_{\text{NS}}^{\mathcal{R}}$, $\alpha_{\Delta\mathcal{R}}$, and $\beta_{\Delta\mathcal{R}}$ are reported in Fig.~\ref{fig:exponential}. The exponents $\beta_{\text{NS}}^{\mathcal{R}}$, $\beta_{\text{NS}}^{\mathcal{R}}$ and $\beta_{\Delta\mathcal{R}}$ are the same in $\Sigma_1^-\cup \Sigma_2^-\cup\ell_{\text{PTL}}$. The exponents $\beta^{\mathcal{R}}_{\text{NS}}$, $\beta^{\mathcal{R}}_{\text{R}}$ and $\beta_{\Delta\mathcal{R}}$ are independent of $h$ in $\Sigma_1^-$, while these exponents depend on (and decreasing function of) both $\gamma$ and $h$ in $\Sigma_2^-$, showing a sharp maximum (cusp) right at line $\ell_{\text{PTL}}$ (Fig.~\ref{fig:exponential}a, ~\ref{fig:exponential}b and ~\ref{fig:exponential}c). $\beta_{\Delta\mathcal{R}}$ is a decreasing function of $\gamma$ and an increasing function of $h$ in $\Sigma^+$ (inset of Fig.~\ref{fig:exponential}c). While $\mathcal{R}_{\text{R}}$ and $\mathcal{R}_{\text{NS}}$ transit between two different exponentially decaying functions on the $\ell_{\text{TRS}}$ line, $\Delta\mathcal{R}$ shows a single decay mode with a well-defined exponent, reported in Fig.~\ref{fig:exponential}d as a function of $\gamma$. The corresponding exponent shows a sharp maximum at $\gamma=1$, i.e. shows an increasing (decreasing) behavior for $\gamma<1$ ($\gamma>1$). The exponents $\beta_X^{U,L}$ (X=NS and R), defined in Eq.~\ref{Eq:ROscillating1} and \ref{Eq:ROscillating2} on $\ell_{\text{TRS}}^\pm$ are shown in Fig.~\ref{fig:exponential}e as functions of $\gamma$. We see that maximal values occur at $\gamma=1$, i.e. the Ising model. Note also that $\beta_X^{U,L}$ tend to zero as $\gamma\to 0$, i.e. XX model, showing that the distribution tends to power-law in this limit. The power-law scaling exponents $\alpha_{\text{X}}^{\mathcal{R}}$ are reported in Fig.~\ref{fig:exponential}f. $\alpha_{\text{R}}^{\mathcal{R}}$ and $\alpha_{\text{NS}}^{\mathcal{R}}$ are unity for all $\gamma$ values, except $\gamma=0$ at which $\alpha_{\text{R}}^{\mathcal{R}}(\gamma=0,h=1)=\alpha_{\text{NS}}^{\mathcal{R}}(\gamma=0,h=1)=2$, and $\gamma=1$ at which $\alpha_{\text{NS}}^{\mathcal{R}}(\gamma=1,h=1)=2$. The ($\gamma,h$) phase diagram for $\mathcal{R}_{\text{R}}$, $\mathcal{R}_{\text{NS}}$ and $\Delta\mathcal{R}$ is shown in Fig.~\ref{fig:schematic}. Figures~\ref{fig:schematic}a and~\ref{fig:schematic}b show the results for R and NS sectors, while Fig.~\ref{fig:schematic}c is for $\Delta \mathcal{R}$.\\

\begin{figure*}[t]
\centering
\includegraphics[width=0.325\linewidth]
{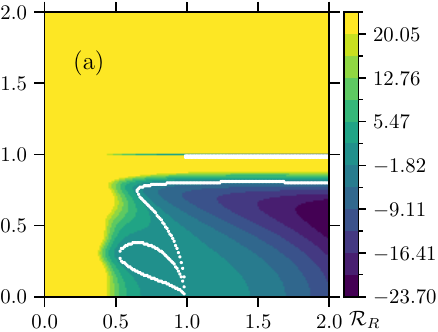}
\includegraphics[width=0.325\linewidth]
{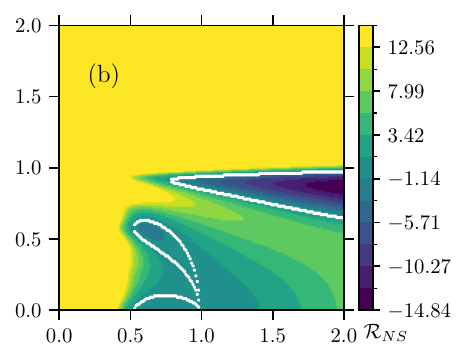}
\includegraphics[width=0.325\linewidth]
{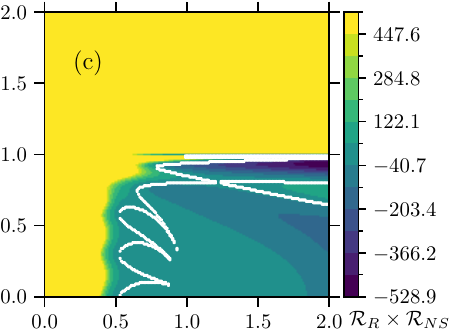}
\includegraphics[width=0.325\linewidth]
{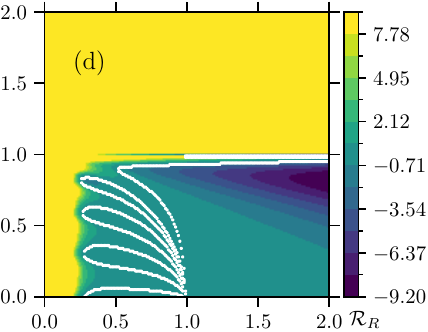}
\includegraphics[width=0.325\linewidth]
{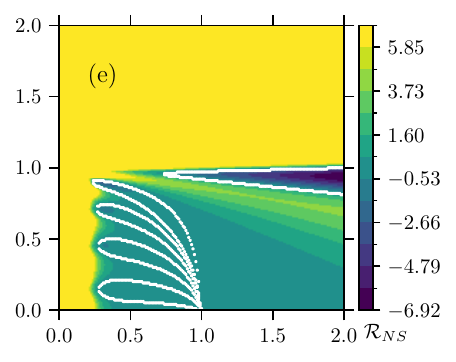}
\includegraphics[width=0.325\linewidth]
{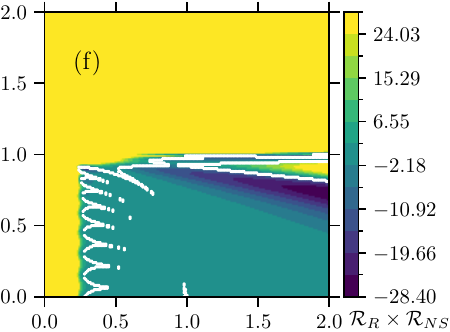}
\includegraphics[width=0.325\linewidth]
{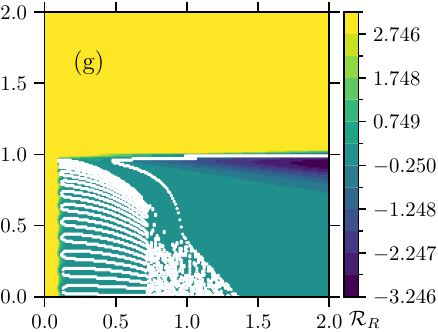}
\includegraphics[width=0.325\linewidth]
{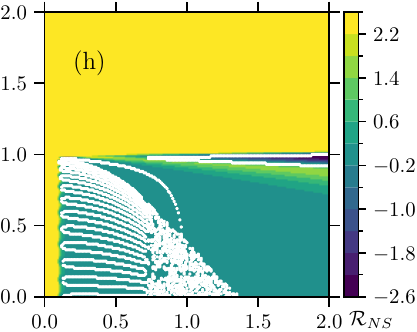}
\includegraphics[width=0.325\linewidth]
{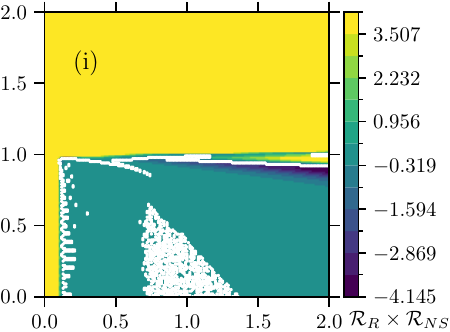}
\includegraphics[width=0.325\linewidth]
{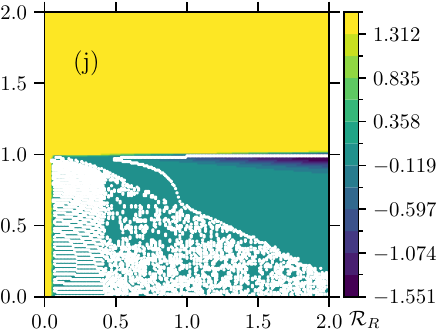}
\includegraphics[width=0.325\linewidth]
{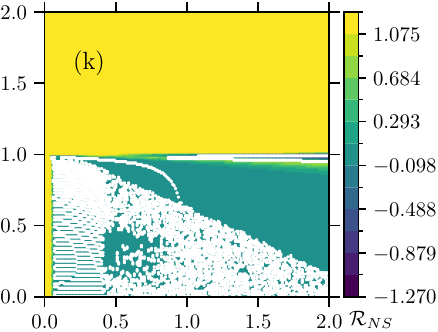}
\includegraphics[width=0.325\linewidth]
{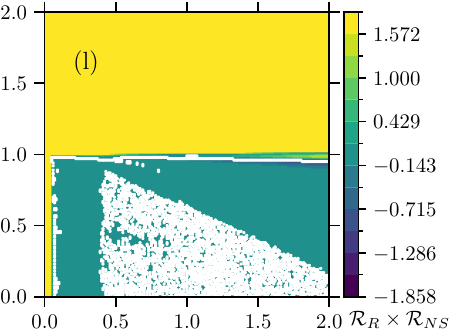}
\caption{Variation of $\mathcal{R}_R$ (first column), $\mathcal{R}_{NS}$ (second column), and $\mathcal{R}_R\times \mathcal{R}_{NS}$ (third column) in the $\gamma-h$ phase space for different values of $L$: $10, 20, 50$ and 100 for first, second, third, and fourth row, respectively. The horizontal axis represents $\gamma$, and the vertical axis corresponds to $h$. To better highlight the regions where the sign of the curvatures change, we’ve limited their range to some $[-\delta, \delta]$, mapping all values greater than $\delta$ (or less than $-\delta$) to $\delta$ (or $-\delta$), respectively. White dots mark the $(\gamma, h)$ points that separate regions with positive and negative values of the curvatures.}
\label{fig:phase_spaces}
\end{figure*}

To gain a better understanding on how the curvature behaves as a function of $L$, $\gamma$ and $h$, we provide a color map phase space representation in Fig.~\ref{fig:phase_spaces}. The colors show the amount of curvature: first and second columns show $\mathcal{R}_{\text{R}}$ and $\mathcal{R}_{\text{NS}}$ respectively, while in the third column $\mathcal{R}_{\text{R}}\mathcal{R}_{\text{NS}}$ is represented for which the negative values show that $\mathcal{R}_{\text{NS}}$ and $\mathcal{R}_{\text{R}}$ are opposite in sign. $L$ increases from row to row as $10$, $20$, $50$ and $100$. The white color (called zero line) shows the regions that the function is zero within the machine size precision, highlighting the boundary between negative and positive values. Focusing on Fig.~\ref{fig:phase_spaces}a ($\mathcal{R}_{\text{R}}$ and $L=10$), we infer that a zero line separates a negative region which matches roughly with $\Sigma_2^-$ from the rest. As $L$ increases, this region matches more better to $\Sigma_2^-$. This corrects the prediction of Eq.~\ref{Eq:RThermodynamicLimit}. More precisely, in the Ramond sector the curvature is negative in $\Sigma_2^-$, while it shows non-trivial behavior in $\Sigma_1^-$. The later is characterized by arcs separating negative and positive regions in much similar way as the arcs in Fig.~\ref{Fig:PhaseDiagramEnergy}, where the ground state of spin chain in interchanged between the ground state of NS and R sectors. The number of arcs increases upon increasing $L$. As $L$ increases further, in addition to the arcs, noisy white regions begin to appear. These regions correspond to areas where the function takes extremely small values, effectively vanishing. Notably, the curvature decreases rapidly with $L$ due to exponential decay, with the fastest decline occurring for $\ell_{\text{PTL}}$, and smaller values of $h$. The most pronounced decay is observed in the $h\to 0$ Ising model ($\gamma=1$).\\

\begin{figure*}[t]
\centering
\includegraphics[width=0.325\linewidth]
{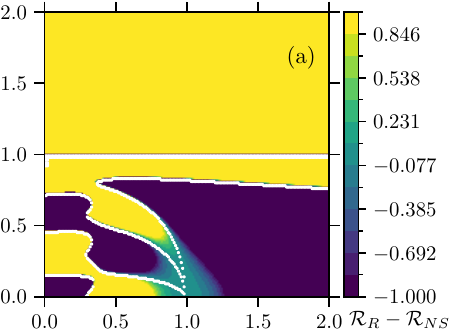}
\includegraphics[width=0.325\linewidth]
{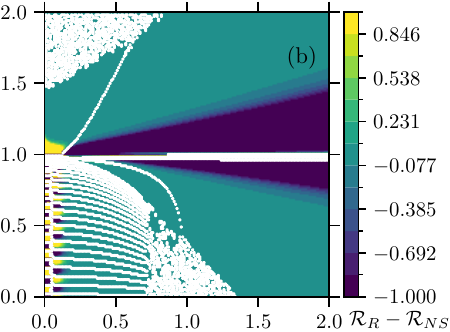}
\includegraphics[width=0.325\linewidth]
{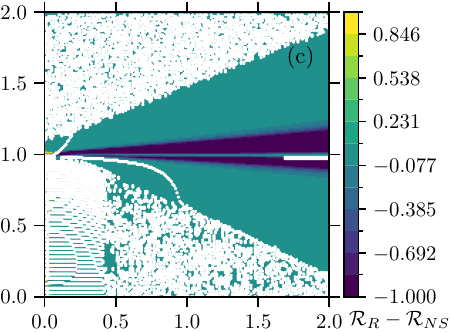}
\caption{Variation of $\Delta \mathcal{R} = \mathcal{R}_R - \mathcal{R}_{NS}$ in the $\gamma$–$h$ phase space for different values of $L$: (a) 10, (b) 50, and (c) 100. The horizontal axis represents $\gamma$, and the vertical axis corresponds to $h$. To better highlight the regions where the sign of $\Delta \mathcal{R}$ changes, we’ve limited its range to $[-1, 1]$, mapping all values greater than 1 (or less than –1) to 1 (or –1), respectively. White dots mark the $(\gamma, h)$ points that separate regions with positive and negative values of $\Delta \mathcal{R}$.}
\label{fig:dR_phase_spaces}
\end{figure*}

For the NS sector, the arcs in region $\Sigma_1^-$ appear in much similar way as the R sector, but the negative region inside $\sigma_2^-$ is quite different. More precisely, the negative region is restricted to a narrow area next to and under the $\ell_{\text{CL}}$ line which narrows upon increasing the system size. Notably, the negative regions in $\Sigma_1^-$ of NS and R sectors are almost complementary.  Therefore we see that the region where the sign of $\mathcal{R}_{\text{R}}\mathcal{R}_{\text{NS}}$ is negative or negligibly small expands to whole $\Sigma_1^-\cup \Sigma_2^-\cup \ell_{\text{PTL}}$ as $L$ becomes larger and larger. Note that, since the absolute value of the difference tends to zero as $L\to\infty$, this difference is not observable in the thermodynamic limit as predicted by Eq.~\ref{Eq:RThermodynamicLimit}. We conclude that the curvature matches the prediction Eq.~\ref{Eq:RThermodynamicLimit} in the R sector, while it exhibits a complementary behavior for the R and the NS sectors in the region $\Sigma_1^-\cup \Sigma_2^-\cup \ell_{\text{PTL}}$. A representation for the arc regions is given in terms of $\Delta \mathcal{R}$ which is shown in Fig.~\ref{fig:dR_phase_spaces} for various $L$ Values, where we see that the zero lines (separators of positive and negative values) enclose regions with arc shape. \\

Given that the change of fermionic number, corresponds to the change in the fermionic parity, consequently leads to a change of the fermionic sector, we infer from the above argument that, the quantum geometry and the Berry curvature depends on the fermionic parity. Putting in another way, the topology of the system changes upon varying the fermionic parity of the system, which is a consequence of the change of the ground state manifold, associated with NS and R sectors. The arcs indicate that within the ordered region $\Sigma_1^-$, there exist distinct subregions distinguished by Berry curvature sign changes, which correspond to boundary condition transitions.

\section{CONCLUDING REMARKS}\label{7}
In this paper, we have explored the ground state and the geometrical aspects of the finite-size XY quantum spin chain, with a particular focus on the Neveu-Schwarz (NS) and Ramond (R) sectors, corresponding to different choices of the boundary conditions of the Jordan-Wigner transformation.

Our analysis reveals notable differences between the NS and R sectors in finite size systems. The energy difference between these two sctors tends to be power-law near the critical points, and exponential in the other regions. Our study shows that boundary conditions play a significant role in shaping the geometric and physical properties of the system, with the NS and R sectors behaving differently in finite chains.

We analyzed the geometrical and topological properties of the system by exploring the ground state manifold which is manifest in the system's curvature. While the NS and R sectors tend to converge in the thermodynamic limit, they show significant difference in finite systems, suggesting that the impact of boundary conditions remains pronounced, leading to distinct behaviors in both sectors. The properties of the Berry curvature in finite system is summarized in Eqs.~\ref{Eq:Rx1},~\ref{Eq:ROscillating1},~\ref{Eq:ROscillating2} and~\ref{Eq:Rx2}, suggesting that $\mathcal{R}_X$ varies either exponentially or in a power-law fashion in terms of $L$, depending on the amount of $\gamma$ and $h$. 

The analysis of the phase-space representation of the Berry curvature reveals that there are some transition lines where the curvature changes sign. We summarize these results in the following:
\begin{itemize}
    \item The lines where the Berry curvature vanishes and changes sign correspond to phase-space separatrices where the system undergoes a topological transition in the fermionic description. The sign change in the Berry curvature suggests a geometric (or topological) transition between different ground-state manifolds. 
    \item These transitions occur as the ground state sector switches between NS and R, meaning the parity of the fermionic vacuum changes.
    \item The increasing number of arcs as the system size grows suggests that these transitions are finite-size precursors of an emerging continuum of sector transitions in the thermodynamic limit. As the system size increases, more regions appear where the curvature sign changes, indicating that the difference between NS and R sectors becomes increasingly sensitive to the precise values of $\gamma$ and $h$.
    \item The NS and R sectors correspond to different fermion number parities, which can be interpreted as different topological configurations of the Jordan-Wigner fermions. The increasing number of curves in the small-$L$ limit suggests an interplay between the bulk topology and boundary-induced effects.
\end{itemize}

Our observations suggests that the vanishing and sign changes of the Berry curvature in the 
$\gamma-h$ space are closely tied to the topological structure of the ground state and the boundary conditions imposed by the Jordan-Wigner transformation. The sector (NS or R) choice depends on whether the total fermion number is even or odd, which in turn depends on the topology of the spin configuration. 

These findings contribute to a deeper understanding of quantum phase transitions and critical behavior in integrable quantum systems. The geometric approach used in this study offers a powerful tool for characterizing complex quantum phenomena, with potential applications in quantum information theory, quantum computing, and the study of quantum critical systems~\cite{nogaki2022even}.\\

\bibliography{refs}

\newpage
\appendix

\section{JORDAN-WIGNER TRANSFORMATION AND FERMIONIC REPRESENTATION OF XY CHAIN} \label{SEC:JW}

In this appendix, we provide a comprehensive derivation of the Jordan-Wigner transformation and its application to the diagonalization of the XY spin chain Hamiltonian with periodic boundary conditions (PBC). This approach allows us to reformulate the spin model in terms of fermionic operators and subsequently express it in a diagonalized form using a Fourier transformation.

\subsection*{1. Jordan-Wigner Transformation}
The XY model Hamiltonian in one dimension is given by:
\begin{equation}
H_{XY} = -J \sum_{l=1}^{L} \left( \frac{1+\gamma}{4} \sigma_l^x \sigma_{l+1}^x + \frac{1-\gamma}{4} \sigma_l^y \sigma_{l+1}^y \right) - \frac{h}{2} \sum_{l=1}^{L} \sigma_l^z.
\end{equation}
To transform this spin system into a fermionic one, we use the Jordan-Wigner mapping:
\begin{equation}
c_l = \left( \prod_{j<l} \sigma_j^z \right) \sigma_l^-, \quad c_l^{\dagger} = \left( \prod_{j<l} \sigma_j^z \right) \sigma_l^+,
\end{equation}
where \( \sigma_l^{\pm} = (\sigma_l^x \pm i \sigma_l^y)/2 \) are spin raising and lowering operators.

Applying this transformation to the Hamiltonian, we obtain:
\begin{equation}
H^{F}_{XY} = \frac{J}{2} \sum_{l=1}^{L-1} \left( c_l^{\dagger} c_{l+1} - c_l c_{l+1}^{\dagger} + \gamma c_l^{\dagger} c_{l+1}^{\dagger} - \gamma c_l c_{l+1} \right) - h \sum_{l=1}^{L} c_l^{\dagger} c_l.
\end{equation}

For periodic boundary conditions, we introduce the transformation:
\begin{equation}
c_{L+1} = -N c_1,
\end{equation}
where \( N \) is the eigenvalue of the parity operator, taking values \( \pm 1 \) depending on the number of fermions in the system.

\subsection*{2. Fourier Transformation and Diagonalization}\label{SEC:App-diagonalization}
We define the Fourier transformation:
\begin{equation}
c_l = \frac{1}{\sqrt{L}} \sum_k e^{i \phi^{(N)}_{k} l} \tilde{c}_k,
\end{equation}
where the allowed momentum modes satisfy \( \phi^{(N)}_{k} = \frac{2\pi}{L} \left( k - \frac{N+1}{4} \right) \) for integer \( k \). Using this transformation, the Hamiltonian simplifies to:
\begin{equation}
\begin{aligned}
H_{XY}^{PBC} &= \frac{J}{2}\sum_{l=1}^L \left(e^{iM} \bar c_{l}^\dagger \bar c_{l+1} + \gamma e^{iM} e^{2iM l} \bar c_{l}^\dagger \bar c_{l+1}^\dagger + \text{H.C.}\right) \\
&\quad - h \sum_{l=1}^L \bar c_{l}^\dagger \bar c_{l},
\end{aligned}
\label{Eq:H_XY-Transformed}
\end{equation}

To diagonalize this Hamiltonian, we introduce the Bogoliubov transformation:
\begin{equation}
b_k = \cos \frac{\theta^{(N)}_{k}}{2} \tilde{c}_k + i \sin \frac{\theta^{(N)}_{k}}{2} \tilde{c}_{-k}^{\dagger},
\end{equation}
where \( \theta^{(N)}_{k} \) is determined by:
\begin{equation}
\tan \theta^{(N)}_{k}= \frac{\gamma J \sin \phi^{(N)}_{k}}{J \cos \phi^{(N)}_{k} - h}.
\end{equation}
The diagonalized Hamiltonian then takes the form:
\begin{equation}
H^{F}_{XY} = \sum_k \epsilon_N(k) \left( b_k^{\dagger} b_k - \frac{1}{2} \right),
\end{equation}
where the excitation spectrum is given by:
\begin{equation}
\epsilon_N(k) = \sqrt{\left(J \cos \phi^{(N)}_{k} - h\right)^2 + \gamma^2 J \sin^2 \phi^{(N)}_{k}}.
\end{equation}

\begin{figure}
\includegraphics[width=0.47\textwidth]{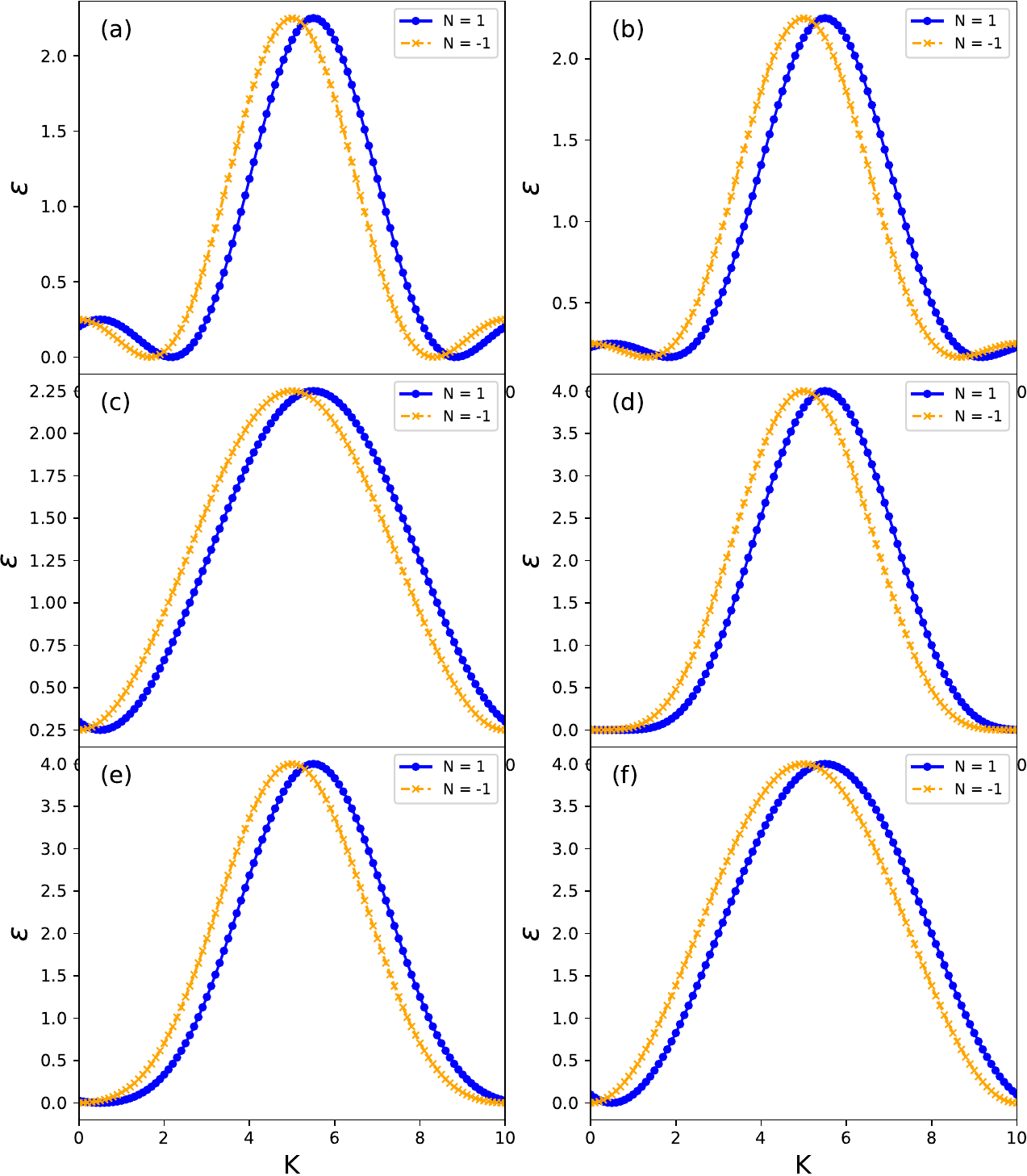}
\caption{$\epsilon_N(k)^2$ for $N = -1$ (R, the line with crosses) and $N = +1$ (NS, the line with dots) and $L = 10$, $J = 1$, $h = 0.5$ and (a) $\gamma=0.0$, (b) 
$\gamma=0.5$, (c) $\gamma=1$. The same graph for $J = 1$, $ h = 1.0$ and (d) $\gamma=0.0$, (e) $\gamma=0.5$, (f) $\gamma=1$. Note that only integer $k$s are permissible. With this in mind, one can decide about the real ground state.} 
\label{Fig:App:R-and-NS}
\end{figure}
Each mode of the Hamiltonian given in Eq.~\ref{Eq:HamilMatrix} can be represented in the isospin space as follows:
\begin{equation}
\begin{split}
\mathcal{H}_k=&\mathcal{H}_k^{11}\left| \uparrow\right\rangle_k\left\langle \uparrow\right|_k+\mathcal{H}_k^{12}\left| \uparrow\right\rangle_k\left\langle \downarrow\right|_k+\\
&\mathcal{H}_k^{21}\left| \downarrow\right\rangle_k\left\langle \uparrow\right|_k+\mathcal{H}_k^{22}\left| \downarrow\right\rangle_k\left\langle \downarrow\right|_k
\end{split} 
\end{equation}
where the isospins are defined as
\begin{equation}
\begin{split}
&\left| \uparrow\right\rangle_k \equiv \left(\begin{matrix}
1\\
0
\end{matrix}\right)_k\equiv {\eta^{(1)}_k}^{\dagger}\left| 0\right\rangle\\
&\left| \downarrow\right\rangle_k \equiv \left(\begin{matrix}
0\\
1
\end{matrix}\right)_k\equiv {\eta^{(2)}_k}^{\dagger}\left| 0\right\rangle.
\end{split}
\end{equation}
Therefore, the ground state given in Eq.~\ref{Eq:GSMain} can be represented as
\begin{equation}
|\text{gs}\rangle = \bigotimes_{k}|\text{gs}\rangle_{k}.
\end{equation}
where 
\begin{equation}
\begin{split}
\left| \text{gs}\right\rangle_k & ={\mathcal{B}_k^{(2)}}^{\dagger}\left| 0\right\rangle \\
&=\left(\left[e^{\frac{i}{2}\theta_k^{N_L}\sigma_x}\right]_{21}{\eta_k^{(1)}}^{\dagger}+\left[e^{\frac{i}{2}\theta_k^{N_L}\sigma_x}\right]_{22}{\eta_k^{(2)}}^{\dagger}\right)\left| 0\right\rangle\\
&=\left[e^{\frac{i}{2}\theta_k^{N_L}\sigma_x}\right]_{21}\left(\begin{matrix}
    1\\
    0
\end{matrix}\right)+\left[e^{\frac{i}{2}\theta_k^{N_L}\sigma_x}\right]_{22}\left(\begin{matrix}
    0\\
    1
\end{matrix}\right).
\end{split}
\end{equation}
Noting that
\begin{equation}
e^{\frac{i}{2}\theta_k^{N_L}\sigma_x}=\left(\begin{matrix}
\cos \frac{\theta^{N_L}_k}{2} & i\sin \frac{\theta^{N_L}_k}{2}\\
i\sin \frac{\theta^{N_L}_k}{2} & \cos \frac{\theta^{N_L}_k}{2},
\end{matrix}\right)
\end{equation}
one finds 
\begin{equation}
|\text{gs}\rangle = \bigotimes_{k}\left(\begin{matrix}
i\sin\frac{\theta^{N_L}_k}{2}\\
\cos\frac{\theta^{N_L}_k}{2}
\end{matrix}\right).
\label{Eq:AppGS}
\end{equation}

\section{EULER-MACLAURIN EXPANSION OF THE ENERGY DIFFERENCE}\label{The EMC expansion theorem}

In this appendix, we provide a detailed explanation of how the Euler-Maclaurin expansion is applied to approximate the energy difference between the Neveu-Schwarz (NS) and Ramond (R) sectors. Our goal is to derive an analytical expression for this energy difference and to examine its behavior in the limit of large \(L\).

The Euler-Maclaurin formula is a powerful tool for approximating discrete sums by integrals along with higher-order correction terms. For a smooth function \(f(x)\) defined over the interval \([a,b]\), the summation formula is given by:
\begin{widetext}
\begin{equation}
\sum_{k=a}^{b} f(k) \approx \int_{a}^{b} f(x)\,dx + \frac{1}{2}\Bigl(f(a)+f(b)\Bigr) + \sum_{n=1}^{\infty} \frac{B_{2n}}{(2n)!}\Bigl(f^{(2n-1)}(b)-f^{(2n-1)}(a)\Bigr),
\label{A1}
\end{equation}
\end{widetext}
where \(B_{2n}\) are the Bernoulli numbers and \(f^{(2n-1)}(x)\) denotes the \((2n-1)\)th derivative of \(f(x)\). This formula indicates that the main contribution of a sum is approximated by its corresponding integral, with correction terms accounting for higher-order refinements.

The energy difference between the NS and R sectors is expressed as sums involving the single-particle energy \(\epsilon_N(k)\). We assume that
\begin{equation}
\epsilon_N(k) = \sqrt{\Bigl(h - \cos\phi_k^{(N)}\Bigr)^2 + \gamma^2 \sin^2\phi_k^{(N)}},
\end{equation}
where the phase is defined by
\begin{equation}
\phi_k^{(N)} = \zeta_k - \frac{\pi (N+1)}{2L}, \quad \text{with} \quad \zeta_k \equiv \frac{2\pi}{L}k.
\end{equation}
Applying the Euler-Maclaurin expansion to the relevant sums, the energy difference between the NS and R sectors is approximated by:
\begin{widetext}
\begin{equation}
E_{NS} - E_{R} \approx \frac{1}{2}\Bigl[\epsilon_{1}(1)+\epsilon_{1}(L)\Bigr] - \frac{1}{2}\Bigl[\epsilon_{-1}(1)+\epsilon_{-1}(L)\Bigr] + \sum_{n=1}^{\infty} \frac{B_{2n}}{(2n)!} \left\{ \Bigl[\epsilon_{1}^{(2n-1)}(L)-\epsilon_{1}^{(2n-1)}(1)\Bigr] - \Bigl[\epsilon_{-1}^{(2n-1)}(L)-\epsilon_{-1}^{(2n-1)}(1)\Bigr] \right\},
\label{Eq:AppB}
\end{equation}
\end{widetext}
where \(\epsilon_{1}(k)\) and \(\epsilon_{-1}(k)\) denote the single-particle energies in the NS and R sectors, respectively, and \(\epsilon_{N}^{(m)}(k)\) represents the \(m\)th derivative with respect to \(k\).

To illustrate the process, we first consider the leading order contribution by approximating the sums with integrals:
\begin{align}
\sum_{k=1}^{L} \epsilon_{1}(k) &\approx \int_{1}^{L} \sqrt{\Bigl(h - \cos\phi_{x}^{(1)}\Bigr)^2 + \gamma^2 \sin^2\phi_{x}^{(1)}}\,dx, \\
\sum_{k=1}^{L} \epsilon_{-1}(k) &\approx \int_{1}^{L} \sqrt{\Bigl(h - \cos\phi_{x}^{(-1)}\Bigr)^2 + \gamma^2 \sin^2\phi_{x}^{(-1)}}\,dx.
\end{align}
In addition, the boundary corrections are given by:
\[
\frac{1}{2}\Bigl[\epsilon_{1}(1)+\epsilon_{1}(L)\Bigr] - \frac{1}{2}\Bigl[\epsilon_{-1}(1)+\epsilon_{-1}(L)\Bigr].
\]
Including first-order corrections from the Euler-Maclaurin expansion (which involve the first derivatives \(\epsilon_{1}'(k)\) and \(\epsilon_{-1}'(k)\)), and after some algebra, we obtain for large \(L\) the following leading contribution:
\begin{equation}
E_{NS} - E_{R} = - \frac{\pi \gamma}{12}\frac{1}{L} - \frac{61\pi^3 (4\gamma^2-3)}{720\gamma}\frac{1}{L^3}.
\label{A3}
\end{equation}

This result consists of two main contributions:
\begin{enumerate}
    \item \textbf{Leading Order Term \( \mathcal{O}(1/L) \):} \newline
    This term originates from the integral approximation and boundary corrections, indicating that the energy difference decreases inversely with the system size \(L\). In the limit \(L\to\infty\), this term vanishes, as expected in conformal field theory and related models.
    
    \item \textbf{Third-Order Correction \( \mathcal{O}(1/L^3) \):} \newline
    This term arises from higher-order derivative corrections in the Euler-Maclaurin expansion. These corrections become significant for smaller \(L\) and refine the approximation beyond the leading order term.
\end{enumerate}
These findings are particularly relevant in the study of Ising-type models, superstring theories, and conformal field theories, where the energy difference between boundary and bulk states plays a crucial role.

\section{THE QUANTUM GEOMETRY TENSOR COMPONENTS} \label{Metric Elements}

In this appendix, we present the detailed calculation of the quantum geometric tensor components \(Q_{\mu\nu}\) for the finite XY spin chain. The quantum geometric tensor is defined as:
\begin{equation}
Q_{\mu\nu} = \langle \partial_{\mu} \psi | \partial_{\nu} \psi \rangle - \langle \partial_{\mu} \psi | \psi \rangle \langle \psi | \partial_{\nu} \psi \rangle,
\label{Eq:QuantumGeometricTensor}
\end{equation}
where \( \mu, \nu \in \{h, \gamma, \phi_k^{(N)}\} \), and \( |\partial \psi \rangle = |\partial \text{gs} \rangle_k \) represents derivatives of the ground state wavefunction.

The parameter \( \phi_k^{(N)} \) is now consistently defined as:
\begin{equation}
\phi_k^{(N)} = \zeta_k - \frac{\pi (N+1)}{2L},
\end{equation}
where \( \zeta_k = \frac{2\pi}{L} k \).

The metric tensor is defined as the real part of the quantum geometric tensor:
\begin{equation}
g_{hh} = \operatorname{Re} Q_{hh}.
\end{equation}

The diagonal component \( Q_{hh} \) is given by:
\begin{equation}
Q_{hh} = \frac{1}{4} \sum_{k} \left( \frac{\partial \theta_k^{(N)}}{\partial h} \right)^2,
\end{equation}
where:
\begin{equation}
\frac{\partial \theta_k^{(N)}}{\partial h} = \frac{-\gamma \sin \phi_k^{(N)}}{(h - \cos \phi_k^{(N)})^2 + \gamma^2 \sin^2 \phi_k^{(N)}}.
\end{equation}
Similarly, \( Q_{\gamma\gamma} \) is given by:
\begin{equation}
Q_{\gamma\gamma} = \frac{1}{4} \sum_{k} \left( \frac{\partial \theta_k^{(N)}}{\partial \gamma} \right)^2,
\end{equation}
where:
\begin{equation}
\frac{\partial \theta_k^{(N)}}{\partial \gamma} = \frac{\sin \phi_k^{(N)} (h - \cos \phi_k^{(N)})}{(h - \cos \phi_k^{(N)})^2 + \gamma^2 \sin^2 \phi_k^{(N)}}.
\end{equation}
The off-diagonal component \( Q_{h\gamma} \) is given by:
\begin{equation}
Q_{h\gamma} = \frac{1}{4} \sum_{k} \frac{\partial \theta_k^{(N)}}{\partial h} \frac{\partial \theta_k^{(N)}}{\partial \gamma},
\end{equation}
which evaluates to:
\begin{equation}
Q_{h\gamma} = \frac{1}{4} \sum_{k} \frac{-\gamma \sin^2 \phi_k^{(N)} (h - \cos \phi_k^{(N)})}{\left( (h - \cos \phi_k^{(N)})^2 + \gamma^2 \sin^2 \phi_k^{(N)} \right)^2}.
\end{equation}

By evaluating these summations numerically or analytically, one can distinguish between critical and non-critical behaviors of the system, where the critical points manifest as singularities in the metric components \( Q_{\mu\nu} \).

\section{ADDITIONAL GRAPHS FOR THE CURVATURE IN PARAMETER SPACE}\label{CURVATURE OF THE PARAMETER SPACE}

\begin{figure*}[t]
\centering
\subfigure[]{\includegraphics[width=0.49\linewidth]
{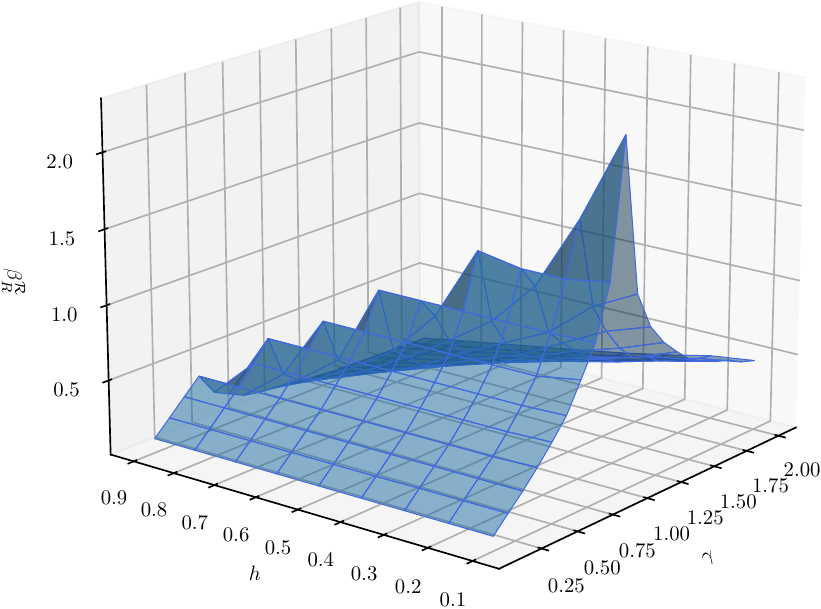}}
\subfigure[]{\includegraphics[width=0.49\linewidth]
{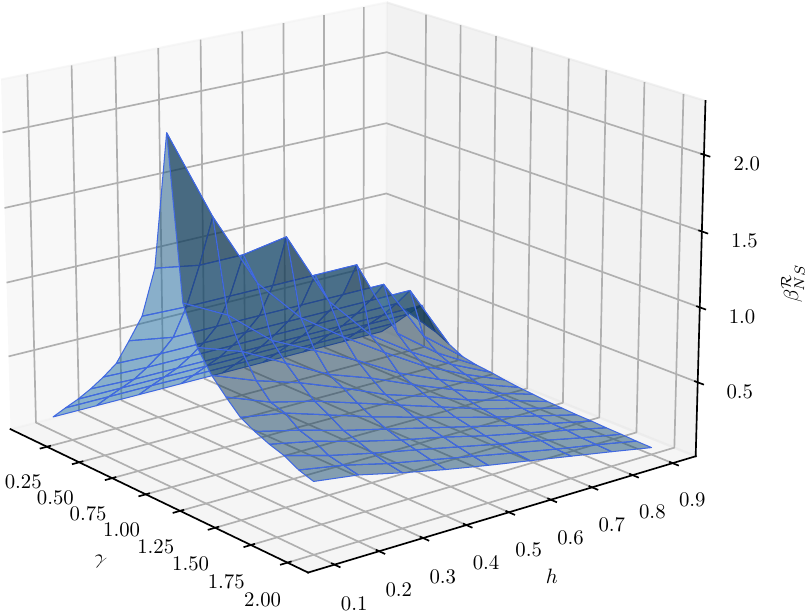}}
\subfigure[]{\includegraphics[width=0.49\linewidth]
{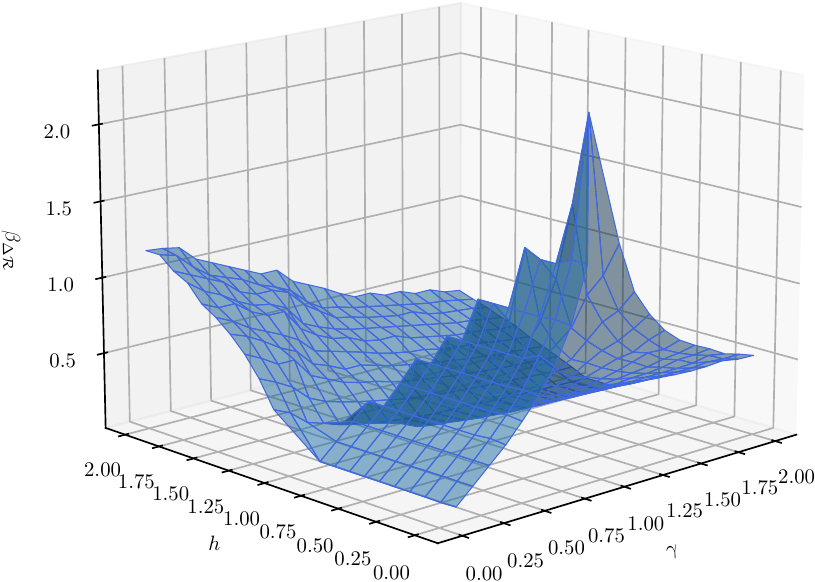}}
\subfigure[]{\includegraphics[width=0.49\linewidth]
{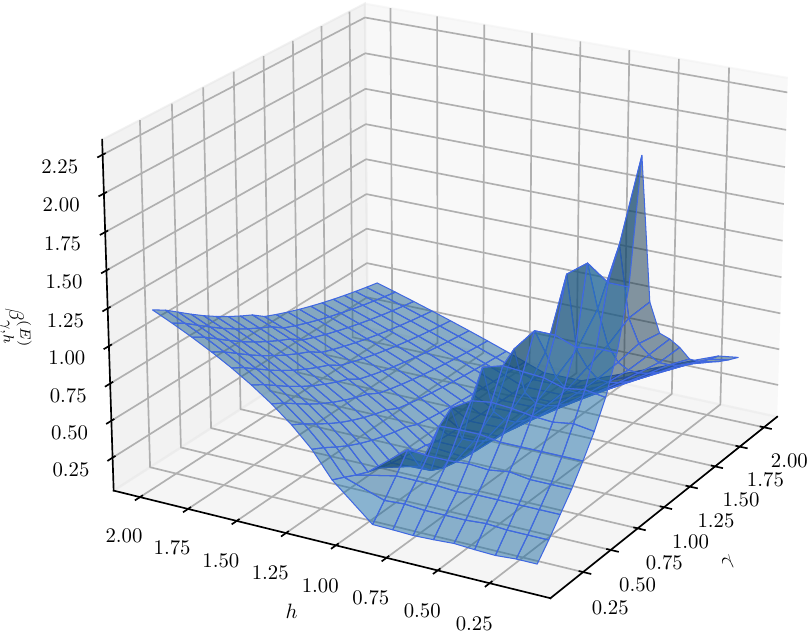}}
\caption{Variation of the exponential exponents of (a) $\beta^{\mathcal{R}}_R$, (b) $\beta^{\mathcal{R}_{NS}}$, (c) $\beta_{\Delta\mathcal{R}}$, and (d) $\beta^{(E)}_{\gamma,h}$ in terms of the allowed values of $\gamma$ and $h$.}
\label{fig:3d_exponential}
\end{figure*}

\begin{figure*}[t]
\centering
{\includegraphics
{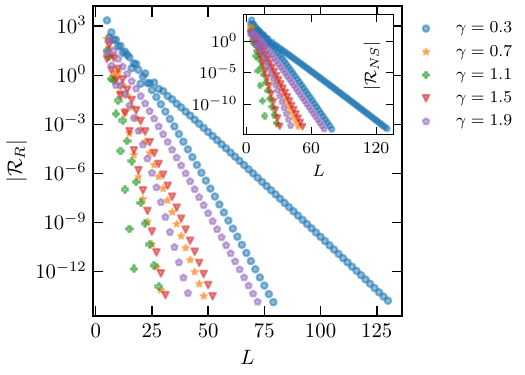}}
{\includegraphics
{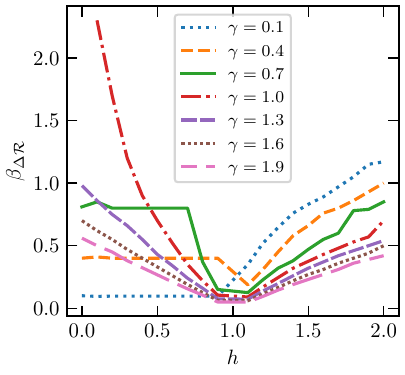}}
\caption{(a) dependence of $|\mathcal{R}_{R}|$ (main panel) and $|\mathcal{R}_{NS}|$ (inset panel) on $L$ for different values of $\gamma$ when $h=0$. (b) dependence of $\beta_{\Delta\mathcal{R}}$ on $h$ for different values of $\gamma$. Note that $h=h_c=1$ is excluded here since the dependence in this case is power-law.}
\label{App:fig:h1}
\end{figure*}

Figure~\ref{fig:3d_exponential} illustrates the dependence of the calculated exponential powers on the parameters \( \gamma \) (anisotropy) and \( h \) (magnetic field) in the exponential regions \( R_1 \) and \( R_2 \). These regions correspond to the cases where the curvature of the system decays exponentially with increasing system size \( L \). The subplots display the variations of the exponential powers for the Ramond (R) and Neveu-Schwarz (NS) sectors (a and b), the absolute difference in curvature between the two sectors \( |\Delta R| \) (c), and \( \alpha \) (d) as functions of \( \gamma \) and \( h \). The results indicate that the exponential powers are highly sensitive to the values of \( \gamma \) and \( h \), with significant differences observed between the R and NS sectors. Additionally, the variation of \( |\Delta R| \) itself follows an exponential behavior. The absence of \( h = 1 \) in subplot (d) is due to the fact that at this value, the curvature exhibits power-law rather than exponential dependence on system size.

Figure~\ref{App:fig:h1} explores the curvature behavior in the special case of \( h=0 \). At this point, the system enters a distinct phase with different behavior from other cases. Subplot (a) displays the magnitude of the curvature in the Ramond \( |R_{\mathcal{R}}| \) and Neveu-Schwarz \( |R_{NS}| \) sectors as a function of system size \( L \), while subplot (b) presents the exponential power \( \beta \Delta R \) as a function of \( \gamma \) at \( h=0 \). These plots reveal that the curvature behavior at \( h=0 \) deviates from other cases and exhibits oscillatory characteristics. Furthermore, the dependence of \( \beta \Delta R \) on \( \gamma \) in this special case is significant, as illustrated in subplot (a), where the curvature behavior for the two sectors differs at \( h=0 \).

\end{document}